\documentclass[%
 aip,
 jmp,%
 amsmath,amssymb,
reprint,%
author-year,%
]{revtex4-1}

\usepackage{graphicx}
\usepackage{dcolumn}
\usepackage{bm}
\usepackage{color}

\def\lapprox{\mathrel{\hbox{\rlap{\hbox{\lower4pt\hbox{$\sim$}}}\hbox{$<$}}}}
\def\gapprox{\mathrel{\hbox{\rlap{\hbox{\lower4pt\hbox{$\sim$}}}\hbox{$>$}}}}

\newcommand{\be}{\begin{equation}}
\newcommand{\ee}{\end{equation}}

\newcommand {\nind} {\noindent}
\newcommand {\mb} {\mathbf}

\newcommand {\bea} {\begin{eqnarray}}
\newcommand {\eea} {\end{eqnarray}}

\begin{document}

\title{Simulations of Emerging Magnetic Flux. II: The formation of Unstable Coronal Flux Ropes and the Initiation of CMEs}

\author{James E. Leake}
\email{jleake@gmu.edu}
\affiliation{College of Science, George Mason University, 4400 University Drive, Fairfax, Virginia 22030.}
\author{Mark G. Linton}%
\affiliation{ U.S. Naval Research Lab 4555 Overlook Ave., SW Washington, DC 20375}
\author{Spiro K. Antiochos}%
\affiliation{ Goddard Space Flight Center}

\date{\today}

\begin{abstract}

We present results from 3D magnetohydrodynamic (MHD) 
simulations of the emergence of a twisted convection
zone flux tube into a pre-existing coronal dipole field.
As in previous simulations, following the partial emergence of the sub-surface flux 
into the corona, a combination of 
vortical motions and internal magnetic reconnection forms a coronal flux rope. Then, in the simulations 
presented here,  external reconnection between the emerging field and the pre-existing dipole coronal field 
allows further expansion of the coronal flux rope into the corona. After sufficient expansion, internal 
reconnection occurs beneath the coronal flux rope axis, and the flux rope erupts up to the top boundary of the simulation domain ($\sim$ 36 Mm above the surface).
We find that the presence of a pre-existing field, orientated in a direction to facilitate reconnection with the emerging field, is vital to the fast rise of the coronal flux rope.
The simulations shown in this paper  are able to self-consistently create many of the surface and coronal 
signatures used by coronal mass ejection (CME) models. These signatures include: surface shearing and 
rotational motions; quadrupolar geometry above the surface; central sheared 
arcades reconnecting with oppositely orientated overlying dipole fields; the formation of coronal flux ropes underlying 
potential coronal field; and internal reconnection which resembles the classical flare reconnection scenario. This 
suggests that proposed mechanisms for the initiation of a CME, such as ``magnetic breakout'', are operating
during the emergence of new active regions.

\end{abstract}

\maketitle

\section{Introduction}

Coronal mass ejections (CMEs) are among the most energetic phenomena associated 
with solar activity and space weather. These giant eruptions of solar  plasma and magnetic 
field are due to the sudden release of energy built up in the complexity of the magnetic 
fields in the solar atmosphere. Many CMEs are associated with {filament channels}, 
where the magnetic fields are strongly sheared, and hence are strongly non-potential and 
have significant free energy \citep{2000JGR...10523153F, Klimchuk2001, LintonM2009}.

Twisted magnetic flux ropes are thought to play a major role in the onset and evolution of CMEs. 
In quiet Sun regions, pre-eruption prominences are interpreted as twisted coronal flux ropes
\cite[e.g.,][]{2010ApJ...724.1133G}. For active region CMEs, line of sight observations are more 
difficult and the scenario is not so clear, hence the use of idealized models is required to fully 
understand the role of flux ropes in these active region CMEs.

Almost all current CME models which include a magnetic field require either a pre-formed 
coronal flux rope  \cite[e.g.,][]{2003ApJ...588L..45R,2005ApJ...630L..97T,2008ApJ...684.1448M}, 
or the formation of a flux rope from a highly sheared active region prior to or during the 
onset of eruption \cite[e.g.,][]{1998ApJ...502L.181A,1999ApJ...510..485A,2000ApJ...529L..49A,
2003ApJ...585.1073A,2008ApJ...683.1192L}. Hence these models rely on 
either the transfer of sheared, non-potential 
field from beneath the surface, or the evolution of potential coronal magnetic field into 
non-potential field by shearing and/or rotational surface motions, as well as flux 
cancellation or magnetic diffusion. Recent simulations of flux emergence have shown that 
the partial emergence of a sub-surface twisted flux tube into the solar atmosphere leads 
to shearing motions \citep{2004ApJ...610..588M} and sunspot vortical motions 
\citep{2009ApJ...697.1529F}, and observations of sunspot rotations have also been 
interpreted as signatures of twisted flux tube emergence \citep{2013SoPh..282..503K}. 
The \textit{ad hoc} surface motions utilized by some CME models are motivated by observations 
of active regions, and the key features of these observations, such as shearing and rotation, 
are most likely a consequence of the emergence of twisted magnetic flux from beneath the surface. 
Hence it can be argued that these CME models rely on the emergence of twisted 
magnetic flux from the convection zone. However they do not self-consistently calculate a
 process for this flux emergence, as they do not include the lower solar atmosphere and 
 convection zone, but instead use boundary conditions which have features that are 
 associated with the emergence of new twisted flux.

Early 3D simulations of flux emergence found that a twisted, buoyant, convection zone magnetic
flux tube only partially emerges, with the 
axis confined to less than ten pressure scale heights (1.5 Mm) above the surface  
\cite[e.g.][]{2001ApJ...554L.111F,2001ApJ...549..608M}.
Later simulations found that a new flux rope structure forms in the corona, and the 
formation mechanism was
attributed to either shearing and rotational motions observed at the surface 
\cite[e.g.,][]{2009ApJ...697.1529F,Leake2013_1}, or magnetic reconnection 
\cite[e.g.,][]{2004ApJ...610..588M,Archontis_Hood_2012}. For simulations 
without any coronal field, this rope expands and rises into the domain with speeds 
up to 33 km/s \citep{2004ApJ...610..588M,2009ApJ...697.1529F}. For simulations 
with a pre-existing straight, constant-strength, field localized in the corona and aligned 
favorably for magnetic reconnection with the emerging field, the confining
 field is removed which allows a faster escape with speeds up to 60 km/s
 \citep{2008A&A...492L..35A,2009A&A...507..995M,2010A&A...514A..56A,Archontis_Hood_2012}. 

In this paper we address the scenario of how an eruptive flux rope can be formed in the corona. We
study the emergence of twisted convection zone magnetic field into the corona and its interaction with a 
pre-existing dipole active region magnetic field, a field that is more complex than the spatially independent
fields used in the previous studies mentioned above.
 We also address the issue of whether dynamical flux emergence of sheared field from the 
 convection zone can capture the signatures and driving conditions used by CME models 
 such as the so-called ``magnetic breakout'' model. This model relies on shear and /or rotational 
 motions to create a sheared arcade from a potential field, and reconnection 
 between different flux systems to initiate an eruption \citep[e.g.,][]{1998ApJ...502L.181A,1999ApJ...510..485A,
 MacNeice04,2008ApJ...683.1192L}. 
 In \citet{2010ApJ...722..550L}, we performed 2.5D simulations of the emergence of 
 twisted magnetic flux from the convection zone into various coronal configurations, 
 such as simple dipole fields and the quadrupolar fields used in the magnetic breakout model.  
 We found that in 2.5D the emergence 
 process is unable to create an unstable configuration in the corona, due to 
 the suppression of the emergence by dense plasma trapped on the emerging field. Further 2.5D simulations 
 by \citet{2013ApJ...764...54L} 
 found that the slippage of magnetic field though the partially ionized regions 
 of the solar atmosphere is a viable mechanism for allowing more magnetic flux to emerge, 
 but still does not create unstable configurations. We therefore concluded that 3D 
 motions are the only remaining plausible mechanism for this paradigm. 
 It was shown by \citet{2009ApJ...697.1529F}, and confirmed in our simulations of
 \citet{Leake2013_1} (hereafter known as Paper I), that during 3D simulations of flux 
 emergence, vortical motions of sunspots, 
 driven by gradients in twist, are capable of twisting the field in the corona, 
 creating a coronal flux rope. Therefore in this paper, we extend the magnetic breakout 
 explorations of \citet{2010ApJ...722..550L} to 3D, and attempt to drive the eruption 
 of a coronal flux rope by emerging a  twisted flux tube into a pre-existing coronal 
 dipole field in a 3D geometry. This pre-existing dipole is designed to represent 
 the decaying field of an old active region. 
 {\color{blue} 
 
 Previously \citet{Roussev_2012} have performed an emergence and eruption simulation 
 in a similar configuration, on a global scale. Current computational resources make the simulation of dynamical emergence and eruption
 on a global scale difficult. Therefore in the study of \citet{Roussev_2012} the surface signature from a simulation of flux emergence into a field-free 
 corona, performed on the same scale as the simulations in this paper, 50 Mm, was used to drive the corona of a global 
 simulation with a dipole field, on the scales of 500 Mm. In order to use the surface data from the 
 small-scale flux emergence simulation to drive the coronal global simulation, three main assumptions were used.
 First, the length scale of the driving data was increased by an order of magnitude to match that of the global simulation.
 Second, the magnitudes of the non-force-free field, plasma pressure, and density, were reduced by 2, 6, and 9 orders of magnitude, respectively, to values representative of the corona, where the magnetic field should be nearly force-free.
 Third, it was assumed that the pre-existing coronal dipole field of the driven simulation did not need to be included in the driving simulation. This simulation produced an eruption-unstable coronal flux rope, which is our goal here, but
the roles that these various non--self-consistent assumptions played in the dynamics is unknown. To explore this
mechanism in detail, in a self-consistent set of simulations, we therefore restrict ourselves to a simulation
length scale of 50 Mm. This allows us to simulate the entire domain in a single simulation.}



 
 

\section{Numerical Method}
\label{Num}
The equations solved, and the domain and boundary conditions used, are exactly the same as used in 
Paper I, and are briefly summarized below.

\begin{figure*}[t]
\begin{center}
\includegraphics[width=\textwidth]{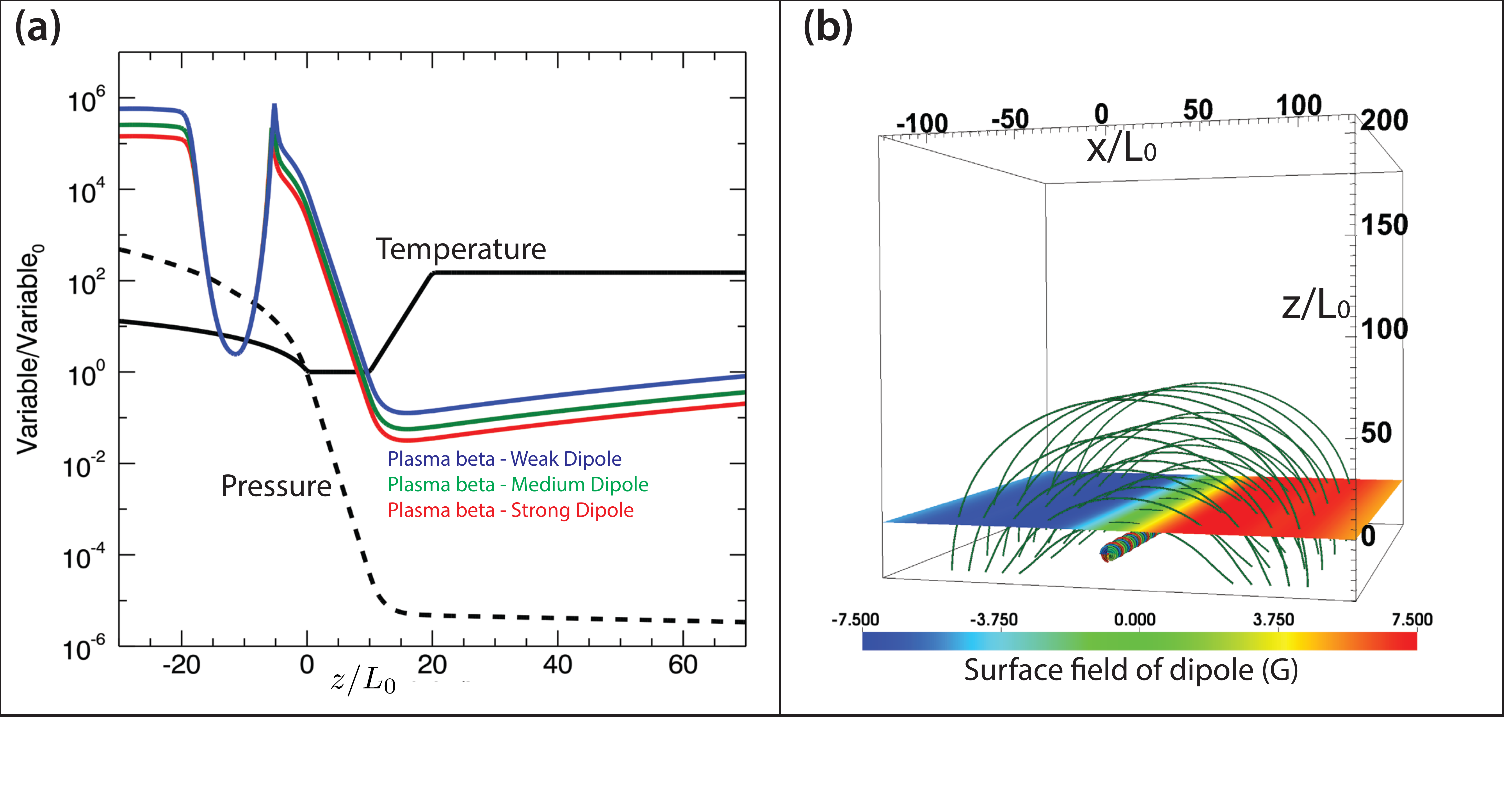}
\vspace{-15mm}
\caption{Panel (a): Initial 1D plasma (temperature and gas pressure) conditions as a function of height for all the simulations in this paper. 
Also shown is $\beta(x=0,y=0,z)$ for the three simulations SD (red), MD (green), and WD (blue). Panel (b): Subset of the domain showing 
selected fieldlines for Simulation MD. The green fieldlines originate from the lower boundary, and belong to the dipole field. The multi-colored
fieldlines originate from the $y=\pm \max{y}$ (side) boundaries and belong to the convection zone flux tube. The transparent surface shows the strength of the vertical field at $z=0$ in G.
 \label{fig:IC}}
\end{center}
\end{figure*}

\subsection{Equations}
The equations are presented here in Lagrangian form:

\begin{eqnarray}
\frac{D\rho}{Dt} & = & -\rho\nabla.\mathbf{v}, \\
\frac{D\mathbf{v}}{Dt} & = & -\frac{1}{\rho}\left[\nabla P 
+ \mathbf{j}\wedge\mathbf{B} + \rho\mathbf{g} + \nabla.\mathcal{S}\right],\\
\frac{D\mathbf{B}}{Dt} & = & (\mathbf{B}.\nabla)\mathbf{v} 
- \mathbf{B}(\nabla .\mathbf{v}) - \nabla \wedge (\eta\mathbf{j}), \\
\frac{D\epsilon}{Dt} & = & \frac{1}{\rho}\left[-P\nabla .\mathbf{v}
 + \varsigma_{ij}S_{ij} + \eta {j}^{2}\right], 
\label{eqn:energy_MHD}
\end{eqnarray}
where $\rho$ is the mass density, $\mathbf{v}$ the velocity, $\mathbf{B}$ the magnetic field, and $\epsilon$ the specific energy density. The current density is given by $\mathbf{j}=\nabla\times\mathbf{B}/\mu_{0}$, $\mu_{0}$ is the permeability of free space, and the resistivity $\eta=14.6 ~ \Omega\textrm{m}$. The gravitational acceleration is denoted by $\mathbf{g}$ and is set to the value of gravity at the mean solar surface ($\mb{g}_{sun} = 274 ~ \textrm{m}\textrm{s}^{-2} \hat{\mb{z}}$). $\mathcal{S}$ is the stress tensor which has components 
$\mathcal{S}_{ij}=\nu(\varsigma_{ij}-\frac{1}{3}\delta_{ij}\nabla.\mathbf{v})$, with
$\varsigma_{ij}=\frac{1}{2}(\frac{\partial v_{i}}{\partial x_{j}}+
\frac{\partial v_{j}}{\partial x_{i}}).$ The viscosity $\nu$ is set to $3.35\times10^{3} ~ \textrm{kg}(\textrm{m}.\textrm{s})^{-1}$, and $\delta_{ij}$ is the Kronecker delta function. The gas pressure, $P$, and the specific internal energy density, $\epsilon$, can be written as
\begin{eqnarray}
p & = & \rho k_{B}T/\mu_{m}, ~ and \\
\epsilon & =  & \frac{k_{B}T}{\mu_{m}(\gamma-1)}  
\label{eqn:eos}
\end{eqnarray}
respectively, where $k_{B}$ is Boltzmann's constant and  $\gamma$ is 5/3. The reduced mass, $\mu_{m}$, is given by $\mu_{m}=m_{f}m_{p}$ where $m_{p}$ is the mass of a proton, and $m_{f}=1.25$. 


\subsection{Normalization}

The equations are non-dimensionalized 
by dividing each variable ($C$) by its normalizing value ($C_{0}$).
The set of equations requires a choice of three normalizing values. We choose normalizing values for the length ($L_{0}=1.7\times10^{5} ~ \textrm{m}$),  magnetic field ($B_{0}=0.13 ~ \textrm{T} $), and gravitational acceleration ($g_{0}=g_{sun}=274  ~ \textrm{m}.\textrm{s}^{-2}$).  From these three values the normalizing values for the density gas pressure  ($P_{0}=B_{0}^2/\mu_{0}= 1.34\times10^{4} ~ \textrm{Pa}$), density ($\rho_{0}=B_{0}^2/(\mu_{0}L_{0}g_{0})=2.9\times10^{-4} ~ \textrm{kg}.\textrm{m}^{-3}$), velocity ($v_{0}=\sqrt{L_{0}g_{0}}=6.82\times10^{3} ~ \textrm{m}.\textrm{s}^{-1}$), time ($t_{0}=\sqrt{L_{0}/g_{0}}$ = 24.9 s), temperature ($T_{0}=m_{p}L_{0}g_{0}/k_{B} = 5.64\times10^{3}~\textrm{K}$), current density ($j_{0}=B_{0}/(\mu_{0}L_{0})=0.609 ~ \textrm{A}.\textrm{m}^{2}$), viscosity ($\nu_{0} = B_{0}^{2}\sqrt{L_{0}/g_{0}}/\mu_{0}=3.35\times10^{5} ~ \textrm{kg}(\textrm{m}.\textrm{s})^{-1}$), and resistivity ($\eta_{0} = \mu_{0}L_{0}^{\frac{3}{2}}g_{0}^{\frac{1}{2}} = 1.46\times10^{3} ~ \Omega$m) can be derived. 
With these values of normalization, and the values of $\nu$ and $\eta$ given above,  the Reynold's number and magnetic Reynolds number in this simulation are both 100.

\subsection{Domain and Boundary Conditions}

The simulation grid is the same as used in Paper I, and is stretched in all three directions. In the vertical direction, $z$, the grid extends from $-30L_{0}$ to $210.45L_{0}$ with a vertical resolution of $0.428L_{0}$ at the bottom boundary and $1.99L_{0}$ at the top boundary. In the horizontal directions, $x$ and $y$, the grid is centered on $0$ and has side boundaries at $\pm 126.85L_{0}$. The horizontal resolution at $x=y=0$ is $0.658L_{0}$, and at the side boundaries is $2.61L_{0}$. 
 As in Paper I, at the boundary all components of the velocity are set to zero, and the gradients of magnetic field, gas density, 
and specific energy density are set to zero.
The resistivity is also smoothly decreased to zero close to the boundary to eliminate diffusion of magnetic field at the boundary. This approach ensures as much as possible that the side boundaries are line-tied. In addition, the velocities are damped close to the horizontal side boundaries and above $z=180L_{0}$ near the top $z$ boundary, as described in Paper I.

\subsection{Initial Conditions}

The initial conditions consist of a hydrostatic background atmosphere which represents the upper $30L_{0}$, or 5.1 Mm, of the solar convection zone, plus the photosphere/chromosphere, and the corona up to $210.45L_{0}$, or 35.8 Mm, above the surface. The temperature gradient in the convection zone is equal to its adiabatic value \citep{STIX}. The photosphere/chromosphere ($0<z<10L_{0}$) is isothermal with temperature $T_{ph}=T_{0}$, and the corona ($z>20L_{0}$) is isothermal with temperature $T_{cor}=150T_{ph}$. There is a transition region between the photosphere/chromosphere and corona ($10L_{0}<z<20L_{0}$) which has a power law profile 
\be
T(z)=\left[\left(\frac{T_{cor}}{T_{ph}}\right)^{(\frac{\frac{z}{L_{0}}-10}{10})}\right]T_{ph}.
\ee

{\color{blue} 
The magnetic field consists of a background dipole field that permeates the entire domain, and a twisted flux tube superimposed in the model convection zone. The dipole field is translationally invariant along $y$, the tube's axial direction, and is given by $\mb{B}=\nabla \times \mb{A}$ where $\mathbf{A} = A_{y}\mathbf{e}_{y}$ and
\be
A_{y}(x,z) = B_{d}\frac{z-z_{d}}{r_{1}^{3}},
\ee
with $r_{1}=\sqrt{x^2+(z-z_{d})^2}$ being the distance from the source. We choose $z_{d}$ to be $-100L_{0}$ so that the initial sub-surface flux tube is far from the source of the 
dipole field. The twisted flux tube is aligned along the $y$ axis, at a height of $z=z_{tube}=-12L_{0}$. The flux tube axial field strength $B_{ax}$ exponentially decays with radius from the center $B_{ax}(r)\sim \exp(-r^{2}/a^{2})$ where $a=2.5 L_{0}$. The tube field has a constant twist $q=-1/a$, with the twist field 
$B_{\theta}(r)= q r B_{ax}(r)$.
}
The tube is perturbed such that it is buoyant at the center ($y=0$) and neutrally buoyant at the ends ($y$ boundaries).
This magnetic configuration is the same as in Paper I, but the dipole field has the opposite orientation ($B_{x}$ for the dipole field has the opposite sign). The background atmosphere and plasma $\beta$ ($2\mu_{0}P/B^2$) along the $z$ axis are shown in Figure \ref{fig:IC}, which also shows a 3D representation of the fieldlines associated with the initial convection zone flux tube and the dipole field.  Above the flux tube axis, the horizontal field $B_{x}$ changes sign at the separatrix between flux tube and dipole, and this separatrix is a favorable location for magnetic reconnection. We perform 4 different simulations with differing strengths of dipole {\color{blue}($B_{d}$)}. We can quantify the  strength of the dipole field relative to that of the flux tube by comparing the azimuthal flux per unit length in ${y}$ in the dipole above the tube
\be
\Phi_{dip} = \int_{z_{sep}}^{z_{top}}B_{x}(x=0,y=0,z)dz,
\ee
to that in the top half of the tube
\be
\Phi_{tube} = \int_{z_{tube}}^{z_{sep}}B_{x}(x=0,y=0,z)dz,
\ee
where $z_{sep}$ is the intersection of the $z$ axis and the separatrix between the tube's field and the dipole field, and $z_{top}$ is the top of the vertical domain.
As the horizontal field $B_{x}$ is nearly antiparallel on either side of this separatrix, under favorable forcing, these two fluxes could reconnect until one of the fluxes is destroyed. However, previous flux emergence simulations show that as the flux tube emerges through the photosphere, only the fieldlines which are concave down and able to shed mass continue to emerge into the corona \cite[e.g.,][]{2001ApJ...549..608M}. 
As in Paper I, we perform three different simulations, Strong Dipole (SD), Medium Dipole (MD), and Weak Dipole (WD), which have decreasing dipole strengths. In this paper, the dipole strengths are chosen such that $\Phi_{dip}/\Phi_{tube}$ = 0.13, 0.1, 0.067, respectively. We also add results from a simulation where no dipole exists (Simulation ND, presented in Paper I).

\begin{figure*}[t]
\begin{center}
\includegraphics[width=\textwidth]{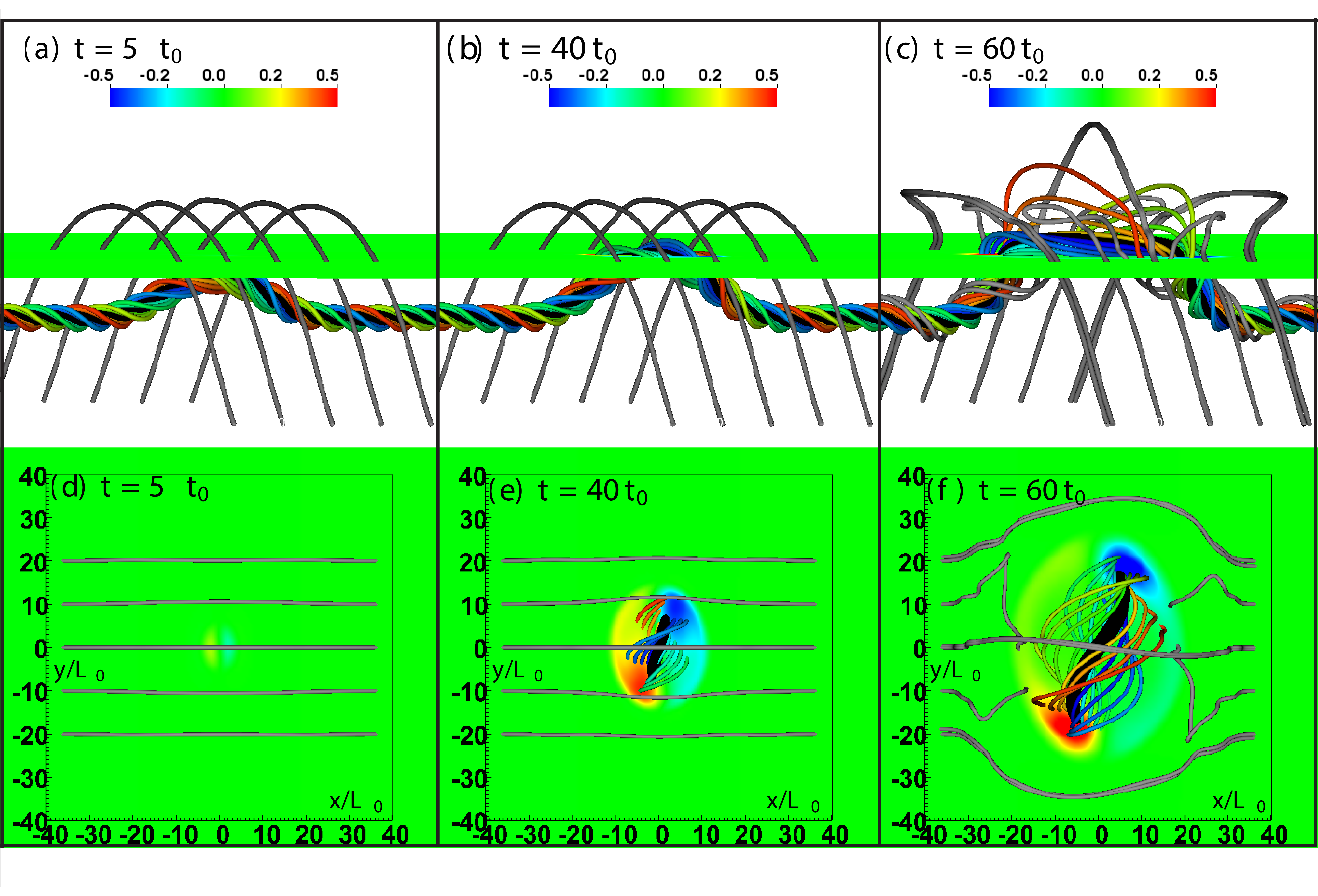}
\vspace{-10mm}
\caption{The early emergence of a convection zone flux tube into a dipole which is orientated opposite to the upper fieldlines of the emerging tube. Panels (a)-(c) show selected fieldlines for simulation MD at times $5t_{0}$, $40t_{0}$, and $60t_{0}$, respectively. The gray fieldlines originate from the lower boundary. The multi-colored fieldlines originate from the side ($y=\pm \max{y}$) boundaries. The black line originates from the axis of the convection zone flux tube
on the side boundary. Each fieldline originates from the same point in each panel. The colored surface shows $B_{z}/B_{0}$ at $z=0$. Panels (d)-(f) show the same times as Panel (a)-(c) but viewed from above. \label{fig:early_emergence}}.
\end{center}
\end{figure*}

\begin{figure*}[t]
\begin{center}
\includegraphics[width=\textwidth]{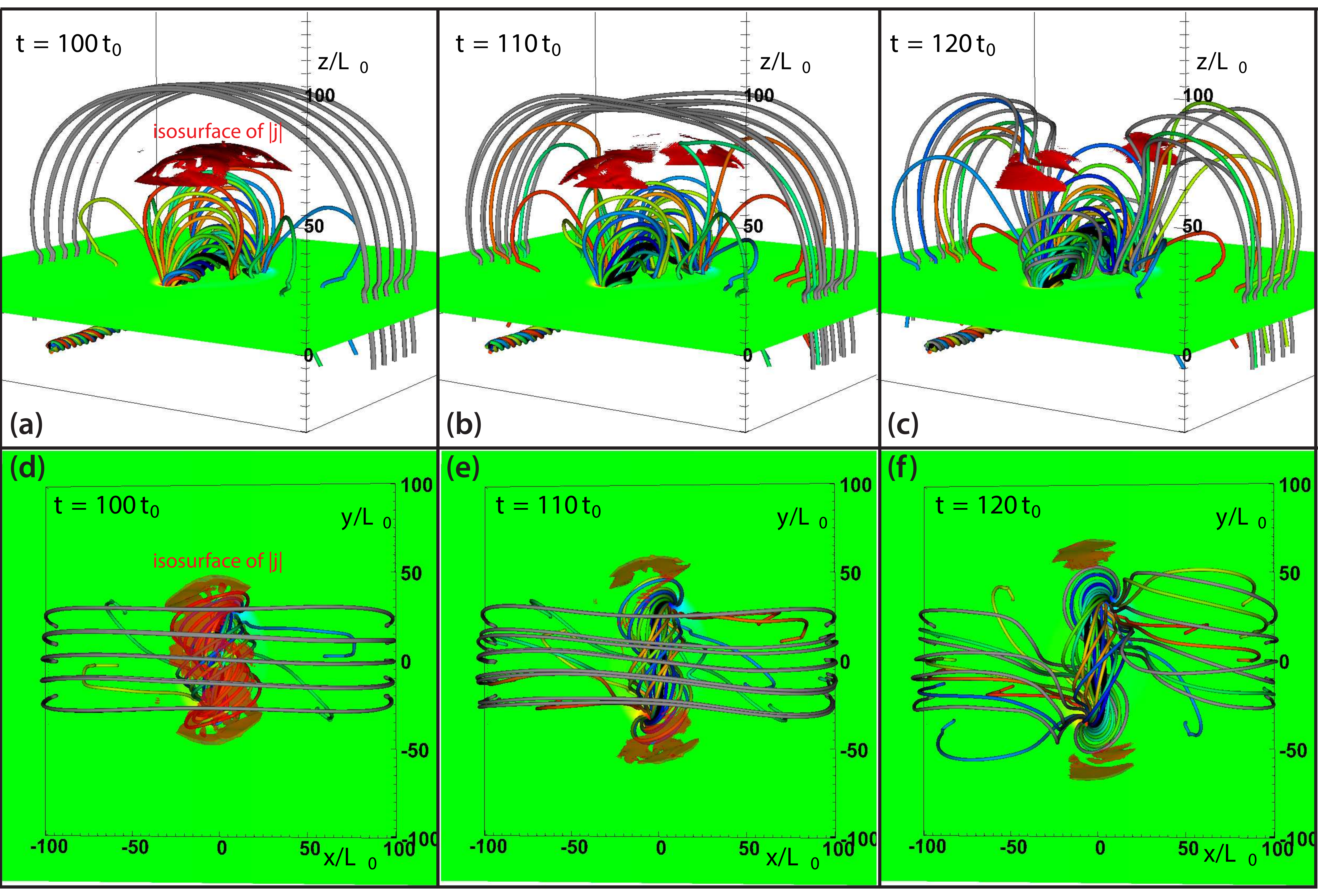}
\vspace{-5mm}
\caption{Formation of a sheared arcade and  external reconnection by flux emergence, in Simulation MD. Selected dipole field (grey lines) and flux tube field (multicolored lines) are shown at $t=100t_{0}$, $110t_{0}$, and $120_{0}$. Also shown are red isosurfaces of $|\mb{j}/j_{0}|>0.004$ above $z=50L_{0}$. 
\label{fig:arcade}}
\end{center}
\end{figure*}

\begin{figure*}[t]
\begin{center}
\includegraphics[width=0.9\textwidth]{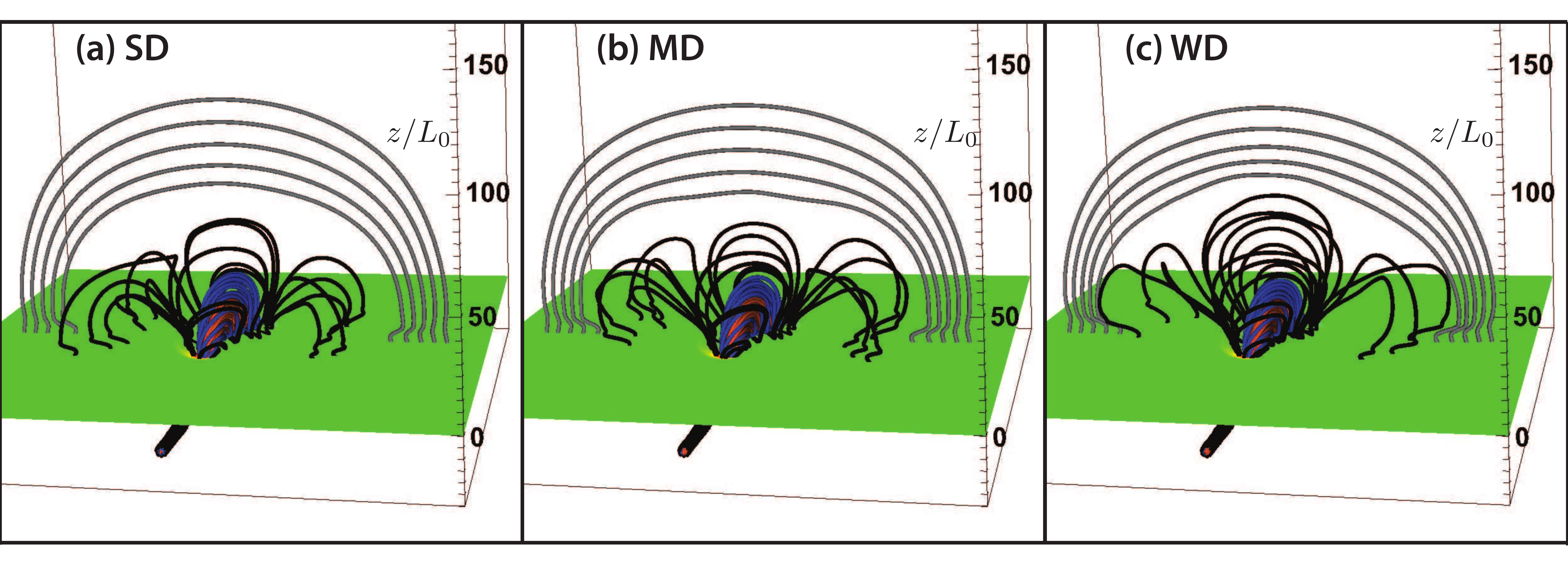}
\vspace{-5mm}
\caption{Selected magnetic fieldlines for Simulations SD, MD, and WD at time $t=100t_{0}$. The gray lines originate from the bottom boundary. The orange, blue and black fieldlines originate from the side boundary, on circles centered on the convection zone flux tube axis at radii of $0.5L_{0}$, $L_{0}$, and $2L_{0}$, respectively. 
\label{fig:compare_dipoles}}
\end{center}
\end{figure*}

\begin{figure*}[t]
\begin{center}
\includegraphics[width=0.9\textwidth]{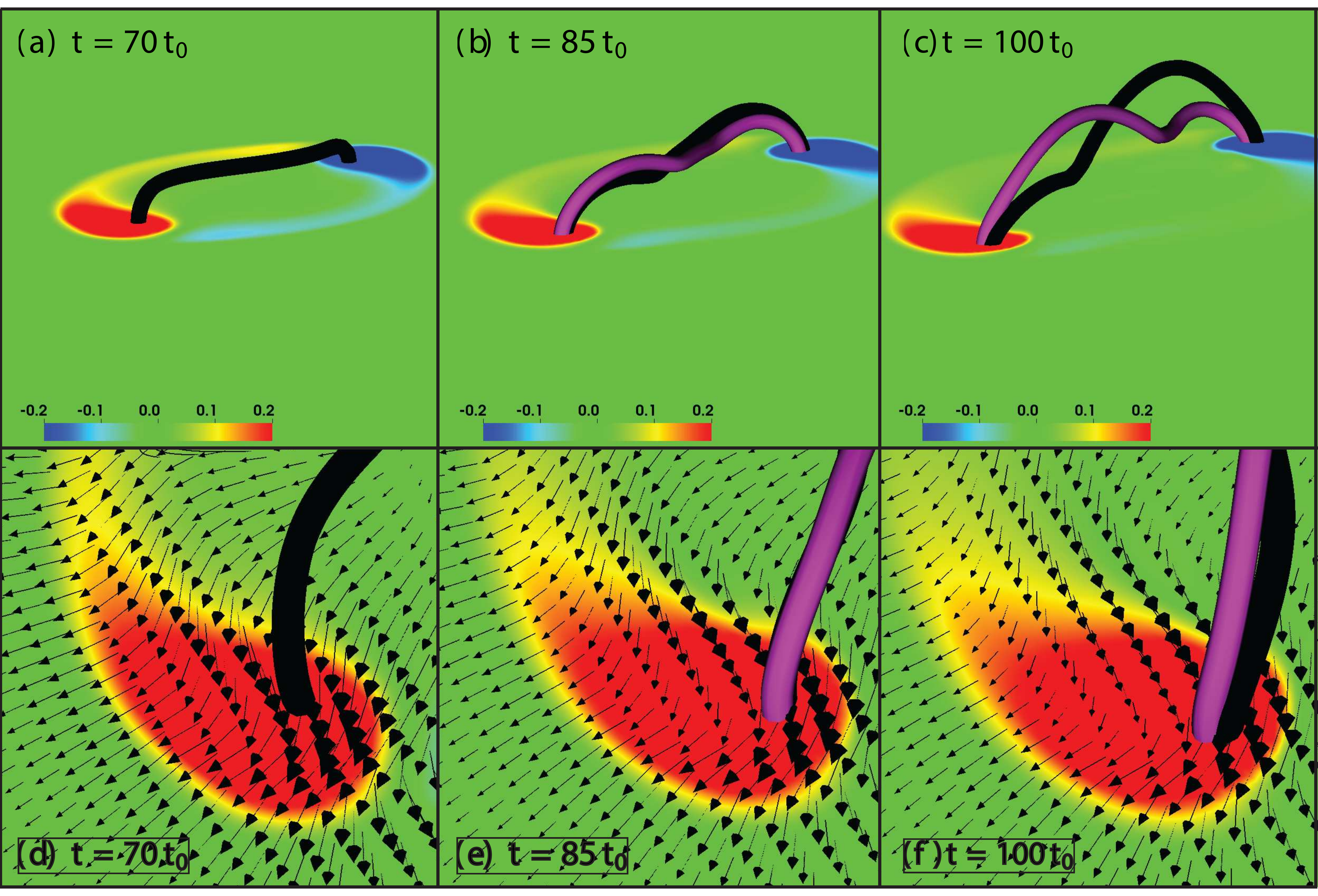}
\vspace{-5mm}
\caption{Rotation of sunspots and twisting of coronal field in Simulation MD. Panels (a)-(c): The colored contour shows $B_{z}/B_{0}$ at $z=0$, the black (purple) fieldline originates form the $y=\max{y}$ ($y=\min{y}$) side boundary. Panels (d)-(f) are a zoomed-in view from above of the positive polarity region (red), with horizontal ($v_{x},v_{y}$) velocity vectors on the $z=0$ surface plane.
\label{fig:rotation}}
\end{center}
\end{figure*}

\begin{figure*}[t]
\begin{center}
\includegraphics[width=0.8\textwidth]{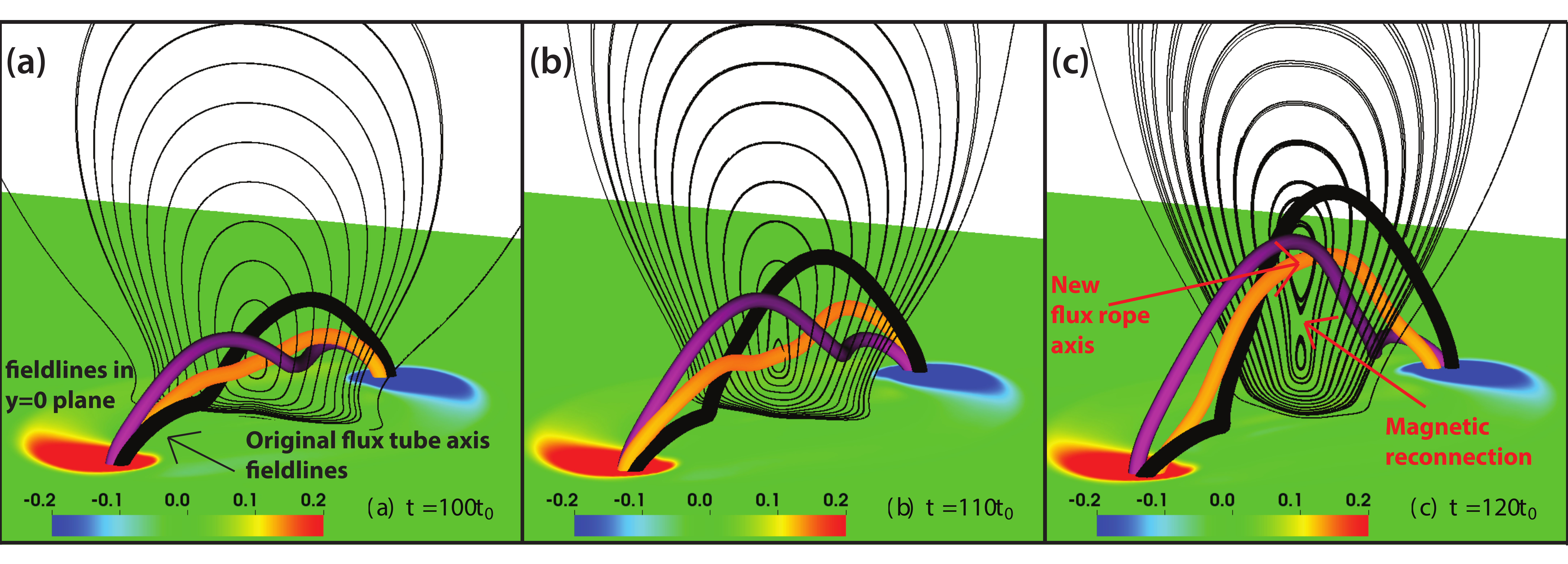}
\caption{Twisting of coronal field and evidence of reconnection in Simulation MD. The black and purple fieldlines are the same as in Figure \ref{fig:rotation}. The thin black lines are $B_{x},B_{z}$ fieldlines in the $y=0$ plane, shown to indicate the location of the O-point in the $y=0$ plane. The orange fieldline in each panel goes through the O-point in the $y=0$ plane (and is not necessarily the same fieldline in each panel).
\label{fig:new_flux_rope}}
\end{center}
\end{figure*}

\begin{figure*}[t]
\begin{center}
\includegraphics[width=0.8\textwidth]{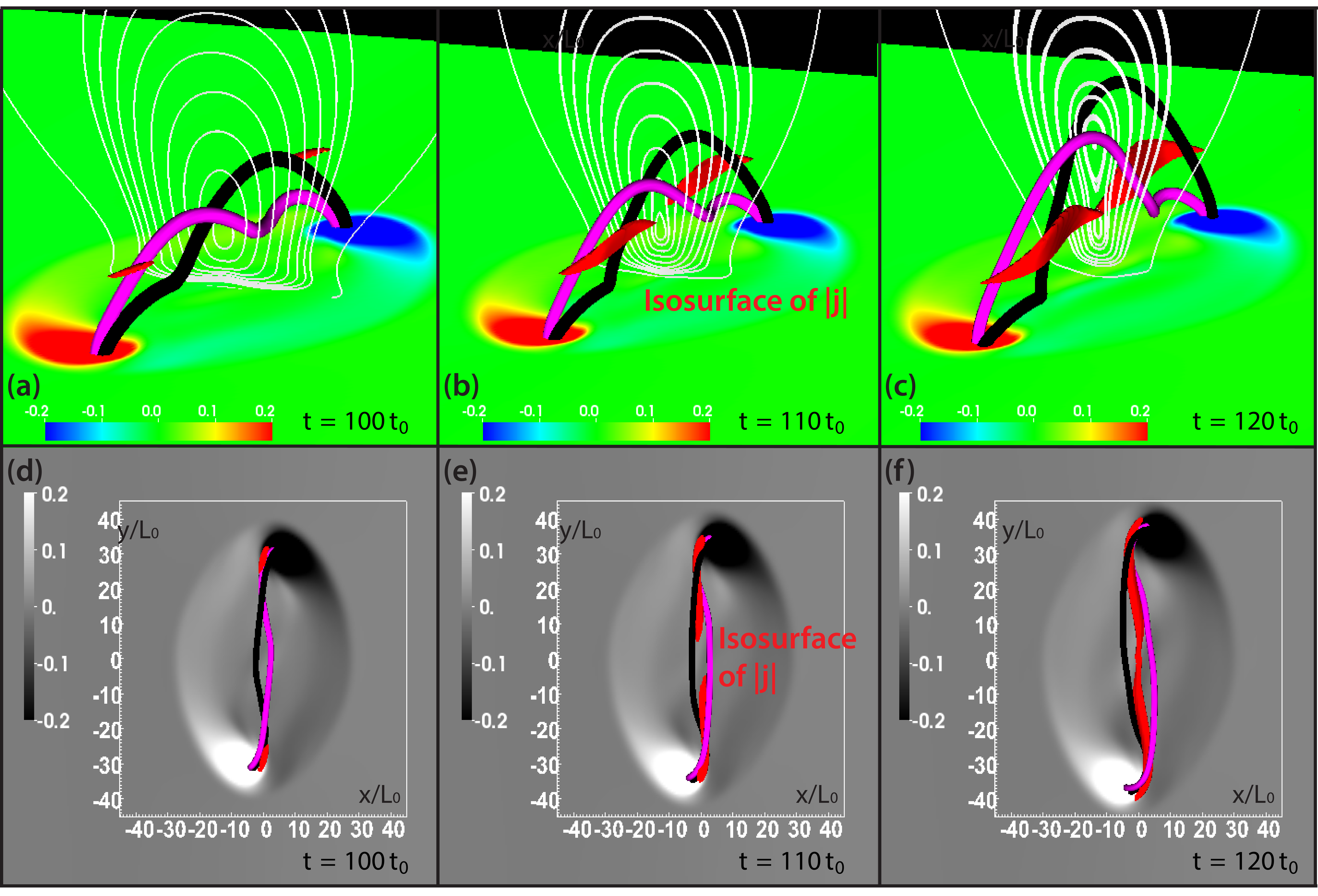}
\caption{Internal reconnection in Simulation MD. The black and purple fieldlines are the same as in Figure \ref{fig:new_flux_rope}. The white lines in the $y=0$ plane are the same as the black lines in Figure \ref{fig:new_flux_rope}. The red isosurface shows $|\mb{j}/j_{0}|>0.2$ in the region above $z=10L_{0}$. Panels (d)-(f) are the same as Panels (a)-(c) but viewed from above and with the color-scale of $B_{z}/B_{0}$ changed to a gray-scale to make the structure of the current sheet clearer.
\label{fig:flare_recon}}
\end{center}
\end{figure*}

\begin{figure*}[t]
\begin{center}
\includegraphics[width=\textwidth]{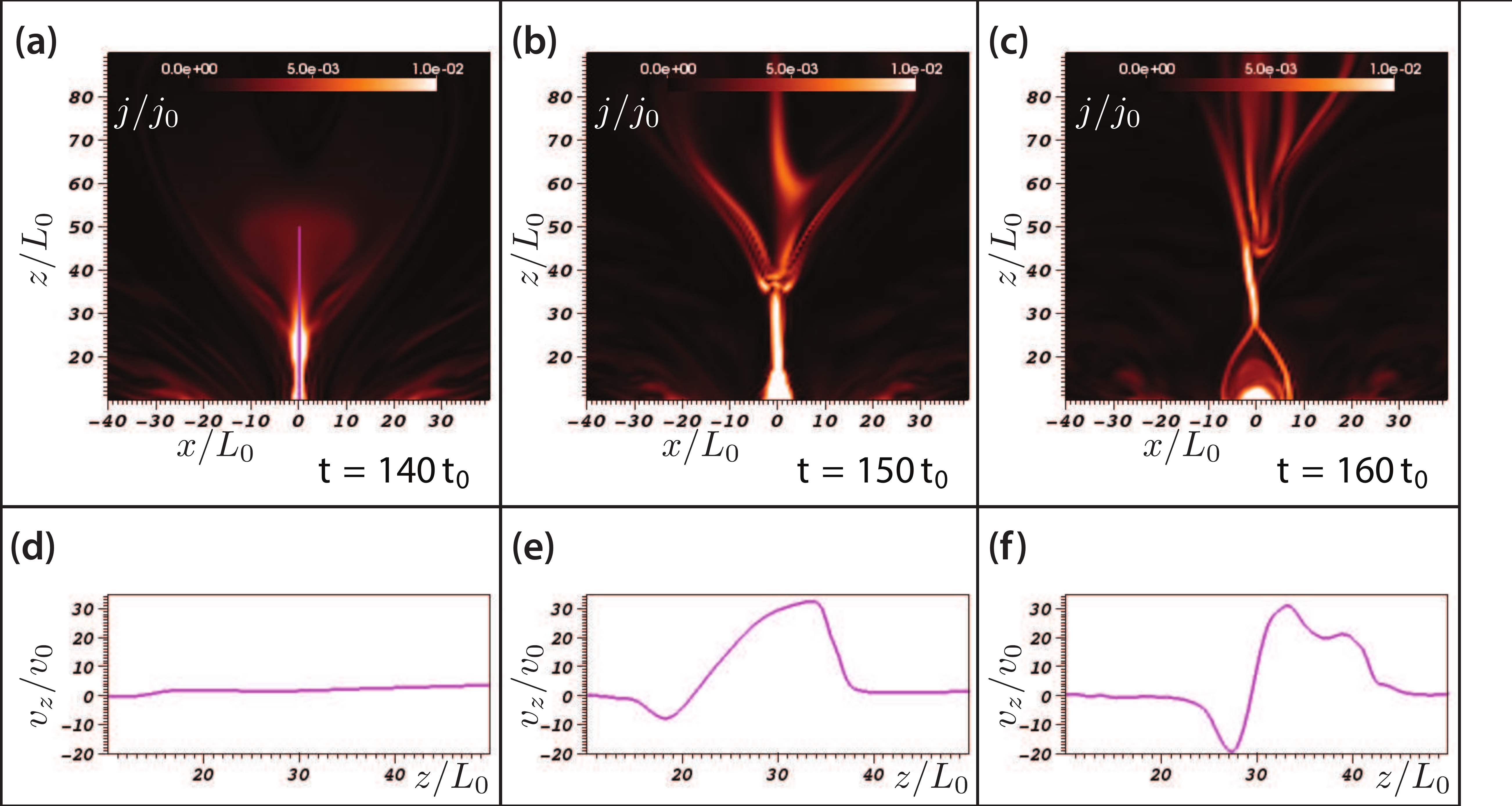}
\caption{Evidence of internal magnetic reconnection in Simulation MD. Panels (a)-(c) show a slice in the $y=0$ plane of the current density $j/j_{0}$ at three different times. A current sheet structure can be seen, beneath the rising flux rope, and after the rope has erupted, the current density structure resembles the standard flare model \citep{Carmichael_1964,Sturrock_1966,
 Hirayama_1974,Kopp_1976}. Panels (d)-(f) show vertical velocity, $v/v_{0}$,
as a function of height along a section of the $x=y=0$ line, indicated by the purple line in Panel (a). Bidirectional vertical flows are observed at later times, indicating reconnection is occurring in the current sheet.
\label{fig:flare_recon_outflows}}
\end{center}
\end{figure*}

\section{Results}
\subsection{Partial Emergence of Convection Zone Flux Tube into 
an Oppositely Orientated Dipole Field: Formation of a Sheared Arcade 
and External Reconnection}

Figure \ref{fig:early_emergence} shows the partial emergence 
of the convection zone flux tube into the overlying dipole field for 
Simulation MD. Despite the orientation of the dipole being opposite 
to that of the dipole in the simulations of Paper I, the early emergence is 
quantitatively similar to the emergence 
in those simulations, as in the convection zone the magnetic field of the 
tube is much larger than that of the dipole.
The sub-surface flux tube rises to the surface, experiences a significant 
horizontal expansion, which is primarily caused by the suppression of the rise by 
 the convectively stable photosphere/chromosphere \citep[e.g.,][]{2004A&A...426.1047A}, 
 and then begins to emerge into the atmosphere via the magnetic Rayleigh-Taylor 
 instability.  Whereas in Paper I the dipole was aligned so reconnection was 
 most favorable beneath the flux tube axis, in the simulations in this paper 
 reconnection is more favorable above the axis, as can be seen in 
 Figure \ref{fig:early_emergence} where some of the gray fieldlines reconnect 
 with the emerging field above the tube's axis. This reconnection of dipole field 
 and emerging field is a continuous process. As the flux of the tube is much 
 larger than the flux of the dipole, reconnection has very little effect on the tube's rise
 in the convection zone. However,  as the tube partially 
 emerges into the corona, the relative amount of flux in the emerging field and 
 dipole field becomes comparable, and magnetic reconnection between the two 
 systems becomes important.

This continued reconnection in the corona between dipole field and emerging field
changes the connectivity of both flux systems. This connectivity change is shown in 
Figure \ref{fig:arcade}, which shows a later stage in the emergence for Simulation 
MD at times $t=100t_{0}$, $110t_{0}$, and $120t_{0}$. The gray dipole fieldlines,
 which originate at the lower boundary, reconnect with the emerging field, and
  leave the domain at the side $y$ boundaries near the axis of the convection 
  zone flux tube. These reconnected fieldlines of the dipole 
  and flux tube create ``lobes'' on either side of the emerging structure, which can 
  be seen in Figure \ref{fig:arcade}, Panels (b) and (c). Figure \ref{fig:arcade} also shows isosurfaces of
     $|\mb{j}/j_{0}|>0.004$ {\color{blue} above $z=50L_{0}$,} where $\mb{j}$ is the current density and 
     $j_{0}=B_{0}/(\mu_{0}L_{0})$. 
       The view from above, in Panels (d)-(f), shows how the connectivity 
      of the field changes and how the structure of the side lobes is formed 
      by the reconnection.  The snapshots shown in Figure \ref{fig:arcade} also highlight 
      how the reconnection acts to remove overlying field and allow further vertical
       expansion of the central emerging structure. Horizontal expansion is restricted by the 
        creation of flux lobes on either side of the arcade

      Figure \ref{fig:arcade}, Panel (b) in particular, shows how this reconnection creates a
      quadrupole structure above the surface, with a central arcade expanding vertically into the corona, an 
  overarching dipole field, and side lobes caused by the reconnection of 
  this central arcade and the dipole field. This is qualitatively similar
   to the magnetic field configuration used in the magnetic breakout CME 
   model \cite[e.g.,][]{2008ApJ...683.1192L} where a quadrupolar field is used 
   as the initial magnetic configuration, with a null point separating the central 
   arcade and the overlying dipole field. 
  In the magnetic breakout CME model, the central arcade is sheared by surface 
 motions to create magnetic field in the axial ($y$) direction, perpendicular to the 
 plane of the arcade (and referred to as ``shear field''). This sheared field 
 drives an outwards expansion, which deforms the original null 
 point between the arcade and dipole field into a current sheet. Reconnection is
 most favorable directly above the central arcade, and reconnection
 at this current sheet above the central arcade, hereafter known as \textit{external reconnection}, 
 allows further vertical expansion. In the simulations in this paper, as in the simulations in Paper I, twist field 
 (with a relatively small component in the $y$ direction) emerges first. 
 Later in time, as the axis of the flux tube 
 emerges through the surface, an increasing amount of magnetic energy is
  present in the shear ($y$) component of the field. Thus the continued 
  emergence of the flux tube creates a sheared arcade structure which is similar to the
   configuration created by shearing motions in the breakout model, as shown
    in  Figure \ref{fig:arcade}.
    
   Figure \ref{fig:compare_dipoles} shows the state of the emergence for Simulations SD, MD, and WD at time 
$t=100t_{0}$. The gray fieldlines show that not much of the dipole field  has
reconnected with the emerging structures. The orange, blue, and black fieldlines originate on circles centered on 
the flux tube axis at the side boundary at radii
of $0.5L_{0}$, $L_{0}$, and $2L_{0}$, respectively.
In all three simulations, we see the same sheared arcade formed, 
with reconnection between emerging field and dipole field creating the quadrupolar structure above the surface. 

The simulations in this paper show that the emergence of a sub-surface flux tube from 
the convection zone into a simple dipole field, orientated so as to favor reconnection 
with the upper twisted fieldlines of the tube, can create the shearing quadrupolar configuration
 used in the magnetic breakout CME model. The axis of the flux rope in this simulation is 
 providing the role of the shear field, and the emergence of the tube from the convection zone 
 brings this shear field into the lower atmosphere, which drives further reconnection 
 between emerging field and dipole field, 
 hence allowing further emergence into the corona. In this way 
 the breakout model has been generalized by being driven by a more realistic emergence of the
 free energy required to drive a CME. We now investigate how the continued emergence 
 process affects this quadrupolar structure, and whether it can create an unstable configuration 
 which erupts.

\subsection{Apparent Rotation of Sunspots}

{\color{blue} As discussed in \citet{2009ApJ...697.1529F} and Paper I, the emergence of the flux tube into
the atmosphere is partial in the sense that only certain portions of fieldlines expand into the corona, while 
other portions remain near the surface. The portions that do emerge expand and lengthen, thus reducing
their twist per unit length, while the non-emerging portions retain the same twist per unit length. This can be
seen in Figures 8 and 9 in Paper I, and in Figure 13 of \citet{2009ApJ...697.1529F}. This gradient in 
twist along fieldlines drives twist to propagation from the convection zone into the corona along the length of the
fieldlines, which results in an apparent rotation of each sunspot about a central point, and the twisting up 
of the emerged coronal field. }


The rotation of the sunspots is represented by the horizontal velocity vectors in Figure \ref{fig:rotation}, 
Panels (d)-(f) at times $t=70t_{0}$, $85t_{0}$, and $100t_{0}$, respectively. In Figure 
\ref{fig:rotation} the black (purple) fieldline originates from the location of the convection
 zone flux tube axis on the $y=\max{y}$ ($y=\min{y}$) boundary. At time $t=70t_{0}$ 
 these two fieldlines are coincident in space. As the twist equilibrates along the fieldlines,
  the sunspots appear to rotate, as indicated by the velocity arrows on the surface ($z=0$), 
 and shown in more detail in Paper I. The two axial fieldlines now diverge due to non-zero 
 resistivity, and appear to wrap around a common point. As in Paper I we designate a 
 new axis by taking the fieldline which intersects the O-point of the in-plane magnetic field
  in the $y=0$ plane, which the black and purple fieldline now twist around. This axis
   fieldline is indicated by the orange fieldline in Figure \ref{fig:new_flux_rope}, which also
    shows the in-plane magnetic field $(B_{x},B_{z})$ in the $y=0$ plane. Early on in
     the emergence ($t \le 70t_{0}$) there is one single O-point above the surface, which 
     is originally intersected by the flux tube axis fieldlines (black and purple lines in 
     Figures \ref{fig:rotation} and \ref{fig:new_flux_rope}). During the emergence, these 
     fieldlines twist around the O-point, and a different fieldline goes through this O-point
     (the orange line in Figure \ref{fig:new_flux_rope}).
     The O-point rises as the central arcade expands vertically due to the external 
     reconnection with the overlying dipole field. At time $t=120t_{0}$, shown in Figure 
     \ref{fig:new_flux_rope}, Panel (c), there are clearly two O-points in the in-plane 
     field above the surface, which indicates internal magnetic reconnection is occurring.
     This will be discussed in more detail in the next section.

      \begin{figure*}[t]
\begin{center}
\includegraphics[width=0.85\textwidth]{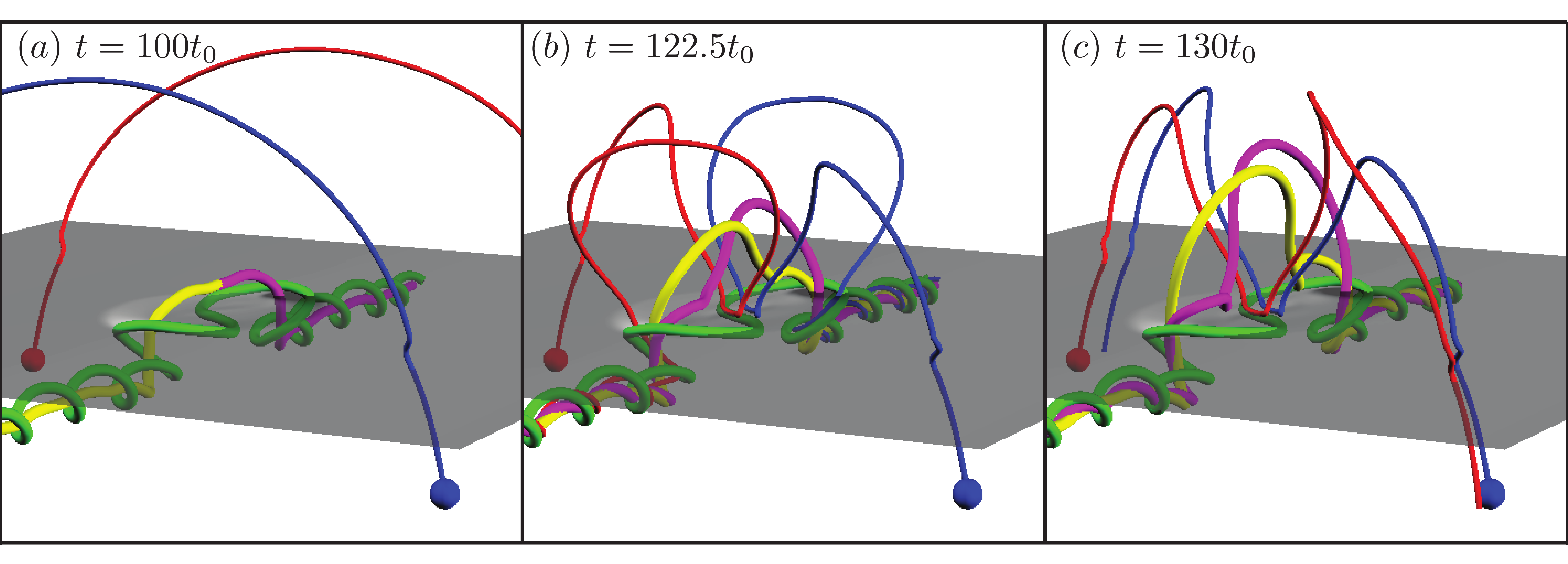}
\caption{Changes in connectivity during the emergence and flux rope formation phases. Panels (a), (b), and (c) show selected fieldlines at times of $t=100t_{0}$, $122.5t_{0}$, and $130t_{0}$, respectively. The red and blue lines originate from the same point in the dipole field at the lower $z$ boundary (the seed locations are denoted by colored spheres). The yellow and purple fieldlines, which coincide in Panel (a), originate on the $y_{min}$ and $y_{max}$ boundaries respectively, again, at the same seed point for each snapshot. The green line is the only fieldline that is not the same in each snapshot (it originates from from the point $(0,0,-1)L_{0})$.
\label{fig:connectivity}}
\end{center}
\end{figure*}
\subsection{Internal Reconnection and Flux Rope Formation}
\label{int_recon}

\begin{figure*}[t]
\begin{center}
\includegraphics[width=0.85\textwidth]{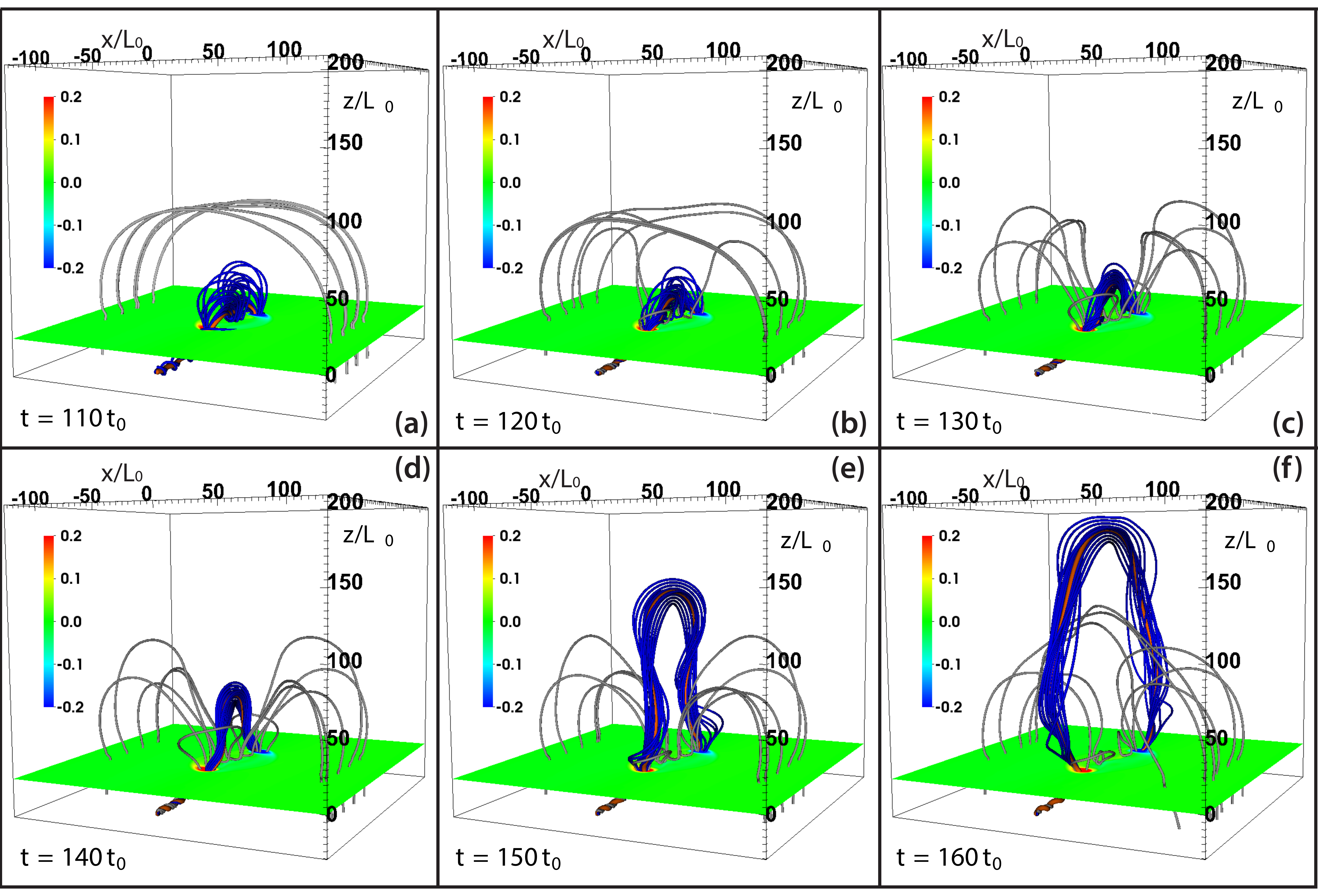}
\vspace{-5mm}
\caption{Eruption of the coronal flux rope in Simulation MD. The horizontal slice shows vertical magnetic field $B_{z}/B_{0}$ on the $z=0$ plane. The grey lines originate on the bottom boundary and represent the dipole field. The orange line intersects the O-point in the $y=0$ plane and is designated the new coronal flux rope's axis. The blue lines are initiated within a circle of radius of $10L_{0}$ from the O-point in the y=0 plane,.
\label{fig:flux_rope_eruption}}
\end{center}
\end{figure*}

Figure \ref{fig:flare_recon} shows the same snapshots in time for Simulation MD 
as Figure \ref{fig:new_flux_rope}, but without the orange fieldline that intersects the 
O-point in the $y=0$ plane. Figure \ref{fig:flare_recon} also shows isosurfaces of $|\mb{j}/j_{0}|>0.2$ 
localized beneath the O-point and above $z=10L_{0}$. Note that the strong currents 
associated with the initial flux tube remain near the surface below $z=10L_{0}$. Panels (a)-(c) in Figure \ref{fig:flare_recon} show a build up in current density above $z=10L_{0}$, 
which is caused by the vertical deformation of the magnetic field. The lines that follow $(B_{x},B_{z}$) in the $y=0$ plane 
indicate that this current is mainly $j_{y} \sim \partial B_{z}/\partial_{x}$ and this is borne out by 
calculations of the individual contributions to the current density. By $t=120t_{0}$ evidence of
 reconnection can be seen, in the form 
of the formation of an X-point in the $y=0$ plane (white lines). 

The current sheet viewed from above in  Panels \ref{fig:flare_recon}(d)-(f) 
resembles two structures which grow and combine in the center of the active region,
 as was also seen in the simulations in Paper I. Various observational studies suggest 
 that the formation and coalescence of J-shaped loops occurs before the formation 
 and eruption of coronal flux ropes \citep{Canfield_1999,Sterling_2000,Liu10}. 
 Whereas in the simulations of Paper I, no obvious evidence of magnetic reconnection, such as
 an X-point or outflows,
 was observed at the site of the current density build up beneath the O-point, X-points 
 are observed in the simulations in this paper. 
 To demonstrate further evidence of reconnection observed in these simulations, Figure \ref{fig:flare_recon_outflows} shows 
 a 2D slice of current density in the $y=0$ plane, at three different times. Along with this slice is a line plot of the vertical velocity along
 the line $x=y=0$. In panel (a), the faint structure above $z=30$ is the flux rope, and the strong vertical structure beneath is
 the current sheet seen in Figure  \ref{fig:flare_recon}, but at a later time. In panel (b) the flux rope has rapidly expanded out of the field of 
 view (this rapid expansion will be discussed in the next section), and the current sheet's extent in the vertical direction has increased.
 Panel (e) shows that there exists a bidirectional vertical flow along the $x=y=0$ line, an indication that reconnection is occurring at a
 height of $z=21L_{0}$. Panels (c) and (f) show that at a later time the site of reconnection has risen up to $z=30L_{0}$, as the flux rope 
 rapidly expands high into the model corona, and ``drags'' the current sheet with it. In panel (f), short loop-like structures below $z=30L_{0}$
 can be seen, and the structure of the current density closely resembles the classical CSHKP flare model \citep{Carmichael_1964,Sturrock_1966,
 Hirayama_1974,Kopp_1976}.
 
 In the simulations of Paper I, the dipole 
 was orientated so as to minimize external reconnection with the emerging structure, 
 and so the vertical expansion of the emerging structure was not as pronounced as in the simulations in 
 this paper. Furthermore, in the simulations of Paper I, there was little direct evidence of internal magnetic reconnection, such as the evidence seen in the simulations in this paper. Internal reconnection is 
 pronounced here due to the strong vertical expansion, facilitated by the external reconnection between 
 emerging field and dipole field. 
 Of the previous flux emergence simulations that exhibit
 evidence of internal reconnection, some had a horizontal coronal field which favored external reconnection and 
 hence  increased vertical
 expansion of the emerging structure \cite[e.g.,][and references therein]{Archontis_Hood_2012}. Others had no dipole field but 
 had a stronger emerging field than that in Paper I \cite[e.g.,][]{2004ApJ...610..588M}, which was able to 
 expand vertically due its own magnetic pressure. Therefore we hypothesize that the likelihood of this
  internal reconnection is related to the extent that the emerging structure can expand 
  vertically to form a strong current sheet beneath the O-point.


As mentioned in Paper I, there have been different proposed mechanisms 
for the formation of a coronal flux rope during the partial emergence of a 
convection zone flux tube. \citet{2009ApJ...697.1529F} suggested that a 
coronal flux rope can be formed when the equilibration of twist along fieldlines 
extending from the convection zone into the corona effectively twists up the
 field in the corona. During this process sections of fieldline which initially
  have a low level of writhe wrap around a new axial fieldline. This process 
  was also observed in  the simulations of Paper I and this paper. On the other 
  hand, \citet{2004ApJ...610..588M}, \citet{2008A&A...492L..35A}, and \citet{Archontis_Hood_2012} 
  suggest that the flux rope is formed by the internal reconnection that occurs 
  when the emerging structure has expanded sufficiently in the vertical direction to 
  create a current sheet. Reconnection at this current sheet converts arched field 
  into twisted field in the corona. As Figure \ref{fig:flare_recon}, Panel (c) shows, it 
  is during this period that two separate O-points can be seen in the $y=0$ plane
   in Simulation MD. One could argue that a precise definition of the formation
   of a new coronal flux rope is when these two O-points can be identified, and that 
   the new flux rope axis is the fieldline which goes through the upper of these O-points.
    This type of internal reconnection can also be seen in the  simulations of 
    \citet{2009ApJ...697.1529F} but was not suggested as a flux rope formation 
    mechanism in that paper. Recent observations of active region CMEs suggest that the formation of a coronal 
       flux rope is associated with a confined flare, which is a consequence of the magnetic 
       reconnection associated with this formation \cite[e.g.,][]{Patsourakos_2013}.
      Despite these findings, it is not clear if
       these two suggested formation mechanisms are independent. Furthermore, we 
       propose that they will both be seen in any simulation such as the simulations in 
       this paper where sufficient vertical expansion of the emerging structure is observed. 

{\color{blue}
Figure \ref{fig:connectivity} highlights the changes in connectivity that lead to a topologically distinct flux rope in the corona. 
 Panels (a), (b), and (c) show selected fieldlines at times of $t=100t_{0}$, $122.5t_{0}$, and $130t_{0}$, respectively. The red and blue lines originate from the same point in the dipole field at the lower $z$ boundary (the seed locations are denoted by colored spheres). The orange and purple fieldlines, which coincide in Panel (a), originate on the $y_{min}$ and $y_{max}$ boundaries respectively, again, at the same seed point for each snapshot. They form part of the erupting flux rope, though they only intersect the O-point in the $y=0$ plane at $t=100t_{0}$, which demonstrates that the axis of the erupting flux rope is not the same throughout the simulations. As the emerging field interacts with the pre-existing dipole field, external reconnection causes fieldlines that were once dipole field to connect at one end to the flux tube, e.g., the red and blue fieldlines in Panel (b). Later in time, in Panel (c), the internal reconnection discussed above can reconnect these blue and red fieldlines so that now they connect from one region of dipole to another, but pass underneath the coronal flux rope, forming an `M' shape. The green line in Figure \ref{fig:connectivity} is the only fieldline that is not the same in each snapshot (it originates from from the point $(0,0,-1)L_{0}$), but is present to show that there is flux tube field that is underneath these M-shaped dipole fieldlines. This the erupting flux rope is topologically distinct to the flux tube field that remains near the surface. 
 The topological separation of coronal flux rope from the original flux tube was originally shown in simulations by \citet{Mactaggart_2014} of the emergence of a toroidal flux rope into a horizontal coronal field. In these simulations, the topological 
 separation happens as a result of external reconnection between emerging field and coronal field, and then internal reconnection which reconnects dipole field underneath the coronal flux rope.}

\subsection{Eruption of a Flux Rope}


Shortly after the start of the internal reconnection observed in Simulations SD, MD and WD, at 
approximately $t=120t_{0}$,  the new coronal 
flux rope rises rapidly into the corona. This is shown for Simulation MD in Figure \ref{fig:flux_rope_eruption}, which 
highlights the opening up of the dipole field by reconnection with the emerging flux structure. 
The simulation is stopped when the erupting flux rope hits the top boundary damping region. 
The blue fieldlines in Figure \ref{fig:flux_rope_eruption} originate inside a circle of radius $10L_{0}$ in the $y=0$ plane, centered on the flux rope axis. Following one of these fieldlines from one polarity region to another, one can see that it completes two turns around the flux rope axis, indicating that the flux rope has significant twist as it erupts.

\begin{figure}[t]
\begin{center}
\includegraphics[width=0.45\textwidth]{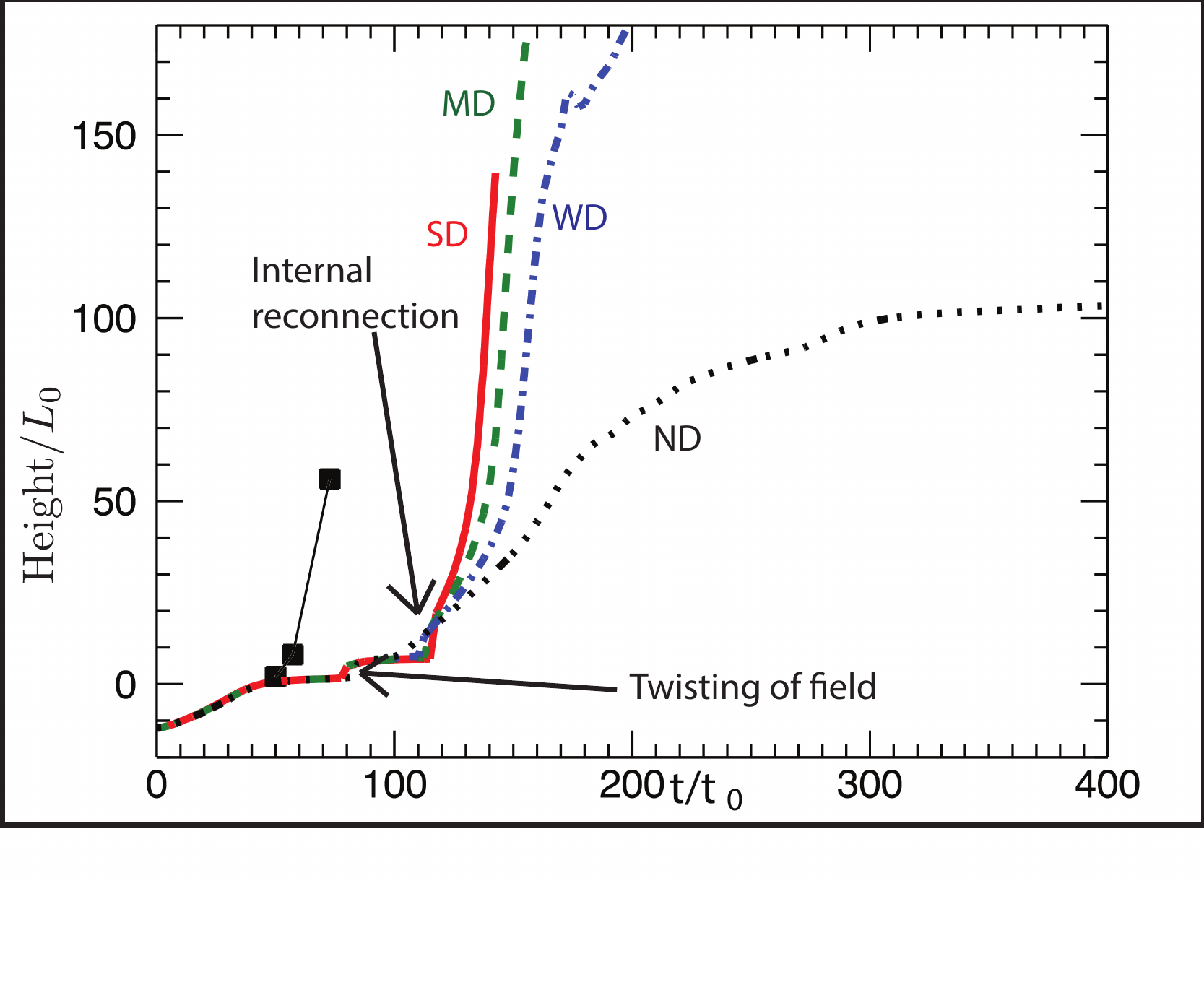}
\vspace{-15mm}
\caption{ Height of the O-point for all four simulations, SD (red solid line), MD (green dashed line), WD (blue dot-dashed line) and ND (black dotted line). Before internal reconnection occurs, i.e., before $t=110t_{0}$, 
the height of the O-point is coincident with the intersection of the convection zone flux tube axis with the $y=0$ plane. After $t=120t_{0}$ there are two O-points, one above and one below the reconnection site, and the coronal flux rope axis is defined by the fieldline which goes through the upper of these two O-points. The top boundary layer begins at $z=180L_{0}$. 
The black squares show the height of the flux rope taken from the simulation of \citet{2004ApJ...610..588M}, where no dipole field was used, but the convection zone flux tube had a stronger magnetic field than in Simulation ND, and was thus more buoyant and able to vertically expand more into the atmosphere.
\label{fig:height-time}}
\end{center}
\end{figure}

Figure \ref{fig:height-time} shows the height of the original flux tube and coronal flux rope's axis 
for all four simulations. To calculate this height at each time, O-points in the $y=0$ plane are
located. For the majority of the 
time in the simulation, there is only one O-point. During the early emergence process, before
 $t=80t_{0}$, this O-point coincides with the intersection of the original convection zone 
 flux tube's axis fieldline with the $y=0$ plane. At $t=80t_{0}$, as a consequence of the twisting 
 of the coronal field mentioned in the previous 
sections, there is a small but rapid jump in the height of the O-point. Following this, in 
Simulations SD, MD, and WD, there is another jump in the O-point at about $t=115t_{0}$.
 This occurs during the period of strongest internal reconnection, 
 as the emerged structure is able to expand vertically 
following external reconnection with the dipole field. During this period, two O-points are
found in the $y=0$ plane, as shown in Figure \ref{fig:flare_recon} and the higher one is 
used to indicate the coronal flux rope's height. For the simulation with no dipole field, ND, there is no such
jump in the height of the O-point, and, as discussed in the previous section, there is little direct evidence for
internal magnetic reconnection. The flux rope slowly rises and the height-time curve indicates that it may be 
asymptoting to an equilibrium.
The three squares in Figure \ref{fig:height-time} show the approximate height of the flux rope shown in Figure 
2 of \citet{2004ApJ...610..588M}, which reports on a simulations with no dipole field  but a stronger magnetic flux tube strength.
As discussed in the previous section, in those simulations, evidence of magnetic reconnection and the rise of the coronal flux rope was reported. 
The height of the flux rope in those simulations reached about $z=55L_{0}$ at about $t=73t_{0}$. In the conclusions of \citet{2004ApJ...610..588M}
it was proposed that the flux rope would not eventually erupt, but be confined by its own field.
As the height of the flux rope in those simulations is less than the height reached in Simulation 
ND ($\sim ~ 100L_{0}$), and as Simulation ND, which also has no dipole field, does not erupt, we agree with their conclusion. Reporting on simulations of the emergence of a flux tube into a field-free corona, \citet{2008A&A...492L..35A} also conclude that the coronal flux rope is confined by its own ambient field. Hence we propose that the presence of the dipole field is vital for the eruption of the coronal flux rope.

\begin{figure}
\begin{center}
\includegraphics[width=0.45\textwidth]{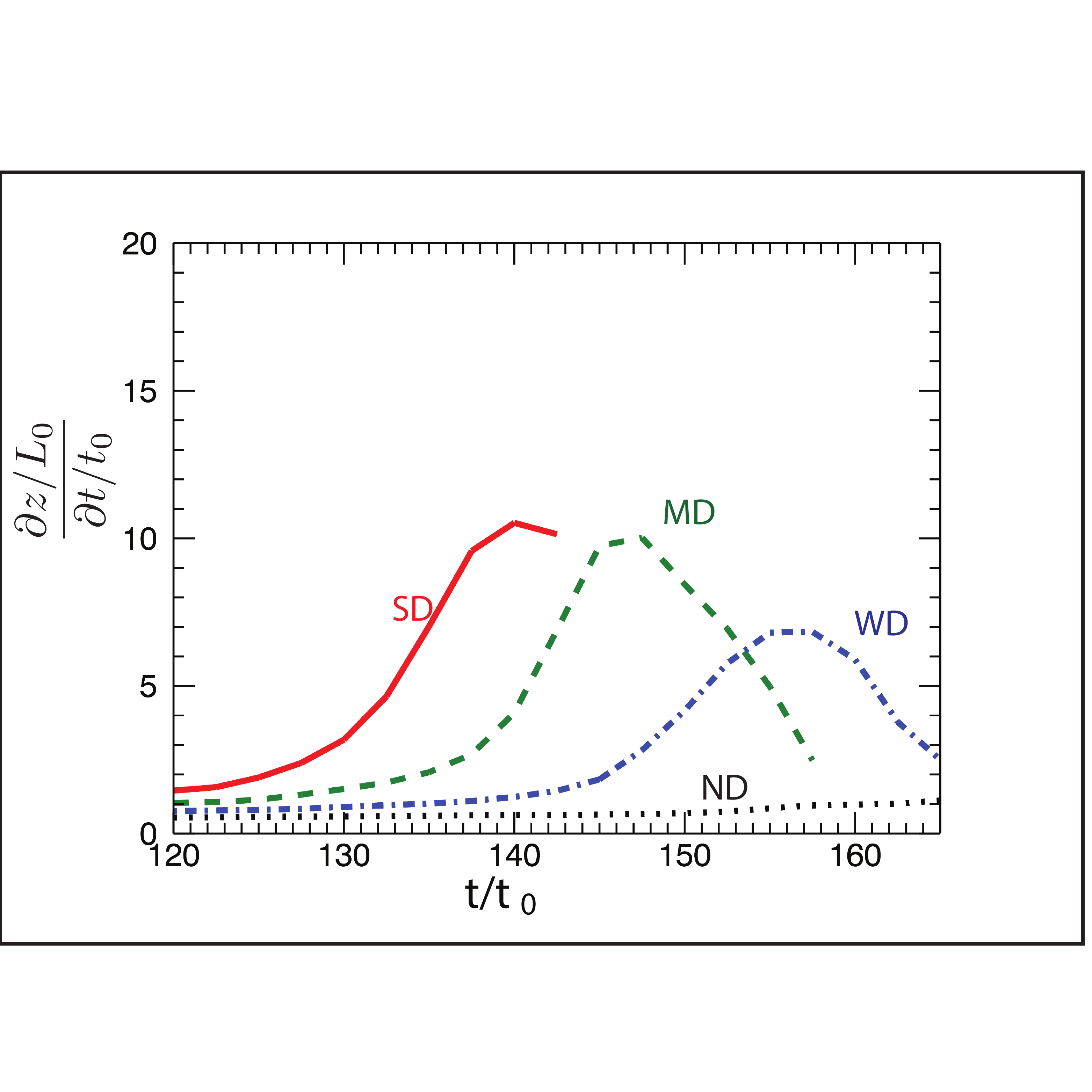}
\vspace{-15mm}
\caption{The vertical O-point velocity $\frac{\partial (z/L_{0})} {\partial (t/t_{0})}$ for Simulations, SD, MD, and WD in the time interval $[120t_{0}:165t_{0}]$, where $z$ is the height of the flux rope. 
\label{fig:rise-rate}}
\end{center}
\end{figure}

After this internal reconnection, for $t>120t_{0}$  the coronal flux rope axis
accelerates. Figure \ref{fig:rise-rate} shows the velocity of the 
flux rope axis for simulations SD, MD, WD and ND in the 
time interval $[120t_{0}:165t_{0}]$.
The peak in the rise speed for Simulation MD is $10v_{0} = 68.2 \textrm{km}/\textrm{s}$.
This is similar to the peak rise speed in the simulations of \citet{Archontis_Hood_2012}, where a horizontal coronal field was used rather than a dipole coronal field.
For all simulations except ND, the speed increases with time, peaks, and then decreases, and the time at which the speed peaks is the time at which the flux rope envelope 
interacts with the upper boundary. Hence we cannot draw conclusions on the further acceleration 
of the flux ropes after this point. Figure \ref{fig:rise-rate} shows that there is an inverse relationship between the peak in the rise speed and the strength of the dipole. Note that as the flux rope in simulation ND appears to asymptote to a certain height at $t=400t_{0}$, the speed of its rise falls to $0.028v_{0} = 190 ~ \textrm{m/s}$.

The strength of the dipole not only affects the maximum rise speed in these simulations, but also the amount of the original convection zone flux tube that reconnects with the
overlying dipole during the emergence and eruption process. The weaker the dipole the less flux reconnects and the larger the resulting erupting coronal flux rope is.
 For Simulation SD, at approximately $t=140t_{0}$, almost all the flux tube has reconnected with the overlying field, and 
it is not possible to identify a coronal flux rope after this time, which is why the red curve in Figures \ref{fig:height-time} and \ref{fig:rise-rate} is halted there. The curves for Simulations MD and WD in 
Figure \ref{fig:height-time} continue until the O-point hits the damping region at $z=180L_{0}$. Because the coronal flux rope in Simulation WD is larger than in Simulation MD, the interaction of the outer 
sections of the rope are affected by the damping region and boundary conditions when the rope axis is at a lower height than in Simulation MD. The decrease in O-point height at $t=175t_{0}$ for Simulation 
WD in Figure \ref{fig:height-time} is a result of this interaction of the flux rope and the damping region and boundary.

\begin{figure}
\begin{center}
\includegraphics[width=0.5\textwidth]{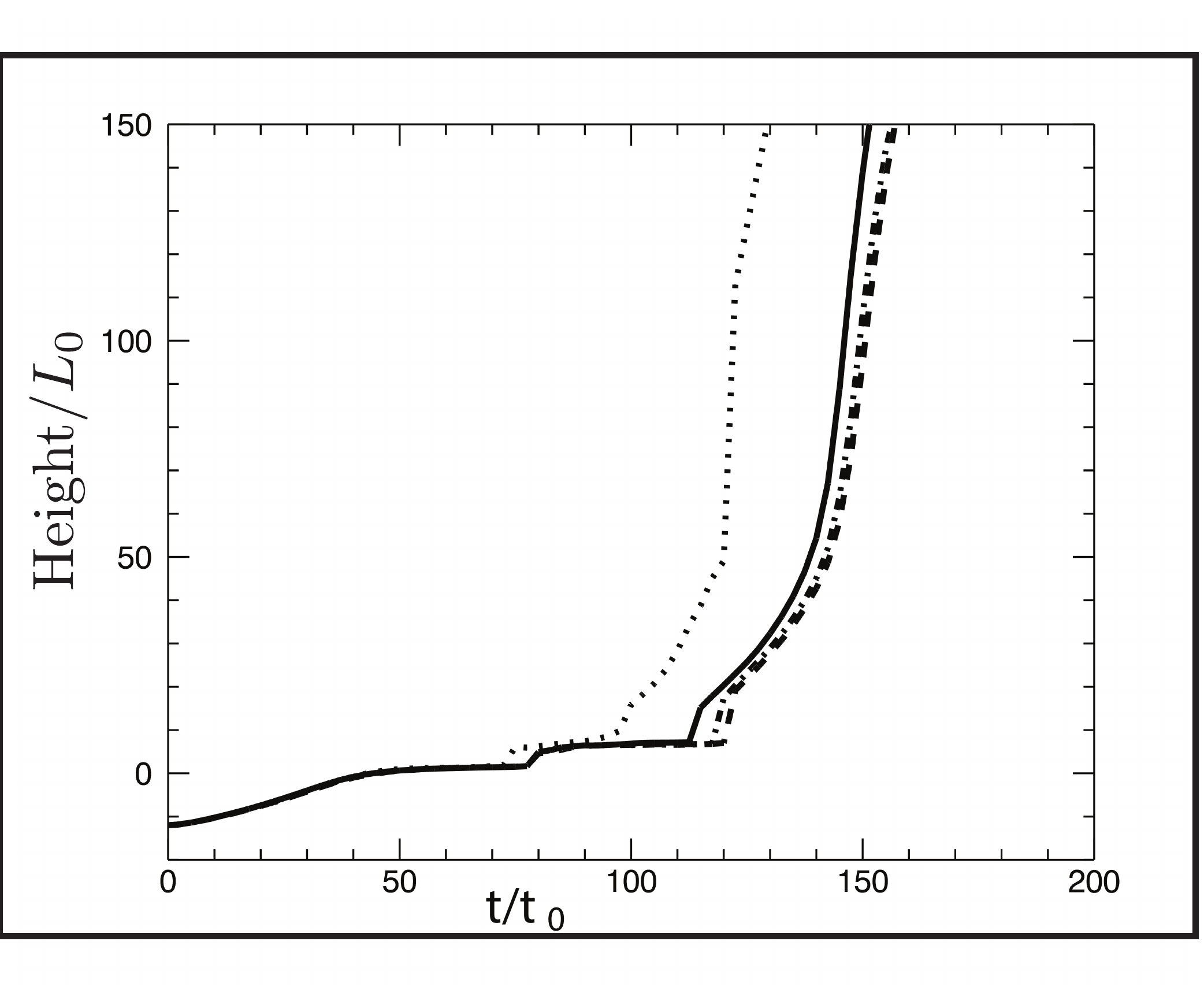}
\vspace{-10mm}
\caption{Effect of resolution and resistivity on eruption of flux rope. Height of the O-point for Simulations MD (grid of $304^3$, solid line), MD2 (grid of $384^3$ dashed line), MD3 (grid of $416^3$dot-dashed line), and MD4 (grid of $304^3$ and $\eta=0$, dotted line).
\label{fig:height-time-convergence}}
\end{center}
\end{figure}

To check for convergence of the solution with resolution, given the choice 
of resistivity in the model, and to test the robustness of these results, we additionally show results for a number of simulations with the same initial conditions as Simulation MD (which has a grid of $304^3$). Simulation MD2 uses a grid of $384^3$ but keeps the same domain and stretching algorithm. Thus the resolution in Simulation MD2 is everywhere $384/304\sim 1.26$ better than Simulation MD. Simulation MD3 uses a grid of $416^3$ so has a resolution $1.36$ better than Simulation MD. Both Simulation MD2 and MD3 use the same numerical parameters ($\eta, \nu$ etc) as Simulation MD. We also present results for  a simulation MD4 which has the same grid as Simulation MD, but where $\eta$ is set to zero. The height-time plot for the coronal flux rope is shown for these additional simulations in Figure \ref{fig:height-time-convergence}, along with that for Simulation MD. Increasing the resolution from Simulation MD (solid curve), through Simulation MD2 (dashed curve), to Simulation MD3 (dot-dashed curve) appears to give convergence, at least in terms of the height of the coronal flux rope. 
The dotted curve  shows the solution for Simulation MD4, with numerical resistivity $\eta/\eta_{0}=0.005$ which exhibits an earlier eruption of the flux rope. 
As in Paper I, we estimate the numerical resistivity using typical Alfv\'{e}n speeds and length scales in regions of strong current and find an effective numerical resistivity of $0.005\eta_{0}$, thus effectively double the Lundquist number of Simulation MD.
We postulate that having a lower effective resistivity in Simulation MD4, compared to Simulation MD, leads to larger current density buildup beneath the flux rope axis and faster internal reconnection, and  
so the required amount of internal reconnection to initiate the eruption is reached earlier in Simulation MD4, compared to Simulation MD.

\begin{figure*}
\begin{center}
\includegraphics[width=\textwidth]{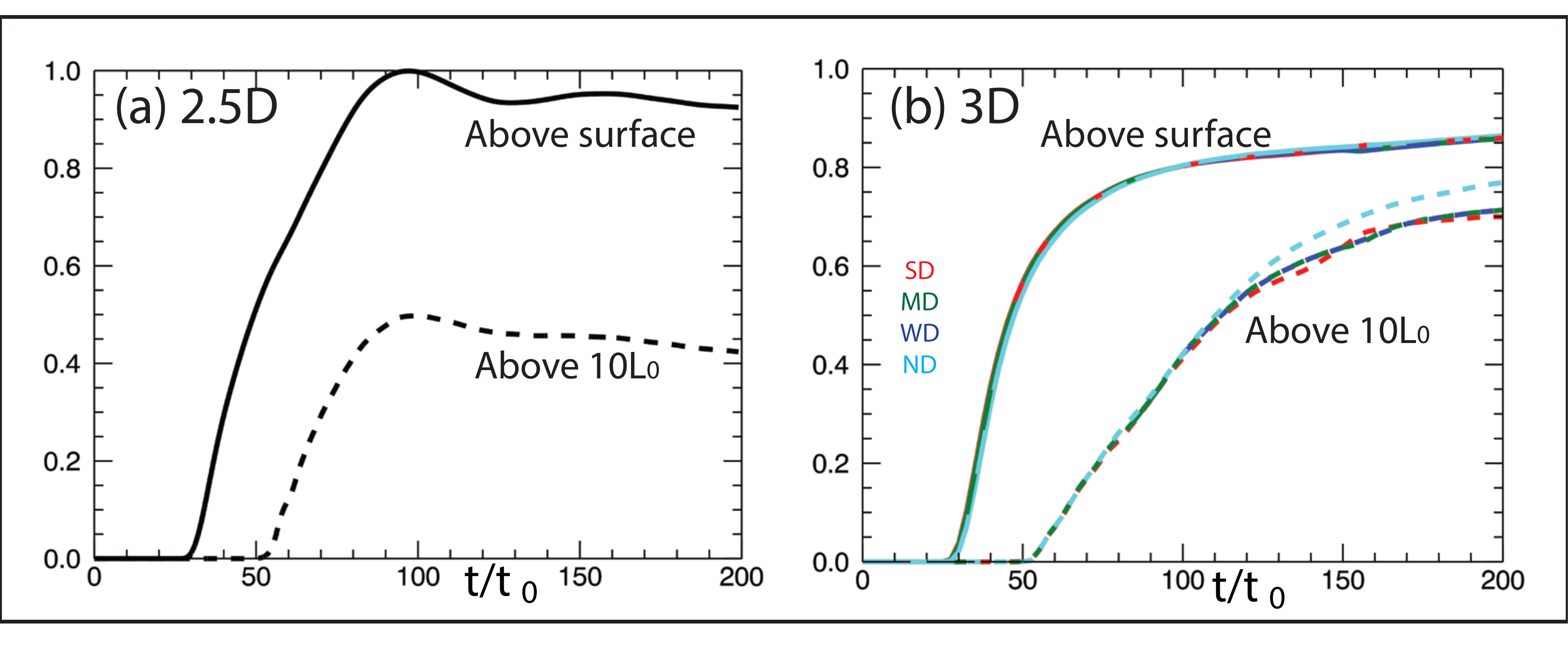}
\vspace{-10mm}
\caption{The ratio of axial flux above a certain height to total axial flux in the domain, for a 2.5D version of Simulation MD in Panel (a), and the full 3D Simulations in this paper in Panel (b) (where the calculation is done in the $y=0$ plane). The two heights chosen are $z=0$ and $z=10L_{0}$. \label{fig:ax_flux}}
\end{center}
\end{figure*}

In \citet{2010ApJ...722..550L} and \citet{2013ApJ...764...54L} we studied flux emergence 
into a dipole in 2.5D, and found that the amount of shear energy supplied to the corona 
was insufficient to drive an eruption. Figure \ref{fig:ax_flux} shows the amount of axial flux
 above a height of $z=0$ and $z=10L_{0}$, normalized to the total axial flux in the domain,
  in the 2.5D simulations of \citet{2010ApJ...722..550L} for a dipole of the same strength as 
  Simulation MD. Also shown is the same calculation in the $y=0$ plane 
  for Simulation MD in this paper. Although in both the 2.5D and 3D  simulations, a large amount 
  of shear flux (almost all in the 2.5D simulation) gets above the surface ($z=0$), only 40\% 
  gets above (and stays above) the $z=10L_{0}$ level in 2.5D, whereas over 60\% achieves this
   in the 3D simulations. This suggests that the eruption requires the
    emergence process to raise sufficient axial flux into the corona to a height at which it is unstable. In 
    the 3D simulations presented in this paper, this is caused by the draining of plasma along fieldlines which allows axial flux to emerge into the corona. 
   Further rise into the corona is driven by   breakout reconnection, which occurs above the coronal flux rope axis and removes overlying dipole field.

\section{Conclusions}

We have performed simulations of the emergence of convection zone flux tubes into the solar atmosphere
with a pre-existing dipole coronal field, focusing on the interaction between emerging flux and dipole field, and
the resulting dynamics. These 
CME simulations  include both the dynamic emergence of magnetic flux into the corona
and a dipole background coronal field representing a line-tied decaying active region. 
This is an improvement of the recent 
simulations of \citet{Archontis_Hood_2012} which included a spatially independent, constant-strength, coronal field. In those simulations, 
the likelihood of eruption increased as the angle between the coronal field and emerging field became more 
favorable to external reconnection. From the results of our Paper I and this paper, we also conclude that the external reconnection is vital to the eruption process, which only happened when the dipole was aligned to maximize 
external reconnection.

Initially the azimuthal flux of the convection zone flux
tube is much larger than that 
of the dipole and so the initial rise and emergence in the low atmosphere (photosphere/chromosphere)
is similar to one where no dipole field exists.
However, as the flux tube only partially emerges, not all its azimuthal 
flux is able to emerge into the corona. At a certain point in the emergence, the azimuthal fluxes of the two flux
systems in the corona (emerging field and dipole field) are sufficiently balanced to make magnetic reconnection important.

The resulting atmospheric magnetic field (above $z=0$) closely resembles the quadrupolar geometry commonly used
in the magnetic breakout model, with a central sheared arcade, side lobes and overlying dipole field. The fact that the
simulations self-consistently produce this configuration without the need for kinematic boundary conditions is a step 
forward in improving the realism of the magnetic breakout model.

External reconnection between the emerging field and the dipole field above the central arcade 
allows further vertical expansion, while the horizontal expansion is suppressed by the formation of the side lobes due to the reconnection above. 
During the continued 
vertical expansion into the corona, equilibration of twist along emerged fieldlines causes apparent sunspot rotations. At present we cannot tell how much 
these sunspot rotation contribute to the accumulation of helicity and magnetic energy in the corona, relative to the other horizontal motions present during the flux emergence 
(e.g., shear flows and siphon flows), and relative to the vertical motions.
This equilibration of 
twist effectively twists up the fieldlines in the corona, distorting the original sheared arcade structure.  
After a certain amount of vertical expansion, there
is a noticeable build up of current density beneath the original flux tube axis and signs of internal (or flare-like) reconnection. 
Two O-points are formed above and below the  X-point of the reconnection site. The former is the intersection of a new coronal flux rope axis and the $y=0$ plane.
We propose that both mechanisms (twisting and reconnection) contribute to the formation of 
a coronal flux rope.

In these simulations, there is no eruption unless there is a pre-existing coronal dipole field, aligned to maximize external reconnection between emerging field and coronal field. Not only 
is the dipole field critical to the likelihood of eruption, but the ratio of dipole flux to convection zone tube flux is also critical to the behaviour of the eruption. The flux rope rise speed, flux rope size, and amount of
reconnection between emerging field and coronal field all depend on the strength of the dipole field relative to the strength of the flux tube field.

Very soon after the internal reconnection rate increases, the new coronal flux rope accelerates into the corona, reaching speeds of the order of 60 km/s, which is consistent with the simulations of \citet{Archontis_Hood_2012}. Further evolution is restricted by the size
of the simulation domain. \citet{Patsourakos_2013} recently observed the formation and eruption of a coronal flux rope, and found that a confined flare occurred immediately prior to the identification of the formation of a flux rope. This rope then rose upwards with speeds of 60 km/s for about 7 hours, before an eruptive flare occurred, with the complete eruption of the coronal flux rope. 
{\color{blue} 
These observations suggest that the simulations in this paper are only capturing the formation and early eruption of a coronal flux rope.
Furthermore, the relatively small domain and fast timescales in these simulations, when compared to an active region CME, mean that further studies with larger domains and active regions are required to see if dynamic flux emergence can capture all the 
stages of a CME eruption, particularly the higher eruption speeds observed in studies such as
\citet{Patsourakos_2013}. }

{\color{blue} We note that the formation of an eruptive flux rope in these simulations is markedly different from that in the simulation of \citet{Roussev_2012}, 
where a similar geometry is used (emerging magnetic field interacting and reconnecting with a coronal dipole). In the simulations of \citet{Roussev_2012} 
a coronal flux rope is formed by reconnection between the emerging field and the coronal dipole (not due to internal reconnection as in the simulations in this paper).
 In \citet{Roussev_2012} null points are formed at either side of the emerging field region, and a plasmoid is formed at the apex of the emerging region, with 
flux from both dipole and emerging structure adding to this plasmoid, as can be seen in Figure 1 of \citet{Roussev_2012}. Because the reconnection which 
forms the flux rope in \citet{Roussev_2012} occurs in a region where little of the axial field of the convection zone flux tube has emerged, the resulting flux rope is 
highly twisted (with $\sim 10$ windings observed). The flux rope in the simulations in this paper is formed by internal reconnection (the external reconnection with 
the dipole field aids the emergence of the sub-surface field into the corona). Thus the reconnection occurs in a region where a significant portion of the axial field of 
the convection zone flux tube has emerged, and so the resulting flux rope has relatively less twist (typically only two windings in the corona are seen). These differences 
between the simulations of \citet{Roussev_2012} and those of this paper are most likely caused by differences in how the flux emergence is treated. In \citet{Roussev_2012}, the emergence at the
lower coronal boundary is driven by surface data from a separate flux emergence simulation which does not include a coronal field, and this data is spatially scaled and 
modified in magnitude to fit coronal conditions. In our simulations the flux emergence and its effect in the corona are both solved self-consistently in a single computational domain. }


These simulations exhibit eruptive behavior of coronal flux ropes \textit{immediately} after formation, which adds some 
evidence to the claim that flux ropes are formed during an eruption, not prior to it. However, the above argument suggests
we may be covering only a portion of a typical active region CME, and so definitive statements on this matter are difficult until
further simulations are performed.

One drawback of successful CME models such as the magnetic breakout model \citep[e.g.][]{2008ApJ...683.1192L,2008ApJ...680..740D} and 
the flux rope CME models which rely on the torus instability 
\citep{2003ApJ...588L..45R,2005ApJ...630L..97T,2008ApJ...684.1448M} is that they do not dynamically 
emerge the sheared magnetic field that is required to initiate the eruption from its origins in the convection 
zone. Typical examples of how this emergence is modeled are via shearing and rotational flows applied 
to the model surface, which creates sheared field and initiates eruptions (in the case of the breakout model), or 
by assuming the coronal flux rope is pre-formed in the corona, or by forming a flux rope via magnetic flux cancellation at 
the surface (in the case of the CME models which rely on the
torus instability). These approaches separate the CME models from the source  
from which they derive their magnetic energy: the convection zone and the solar dynamo.
 To better predict space weather, one important step is to eliminate this separation.
The simulations shown in this paper, which include simple improvements on previous works, such as
the use of a dipole field in the corona, are able to self-consistently create many of the surface and coronal 
signatures used by some CME models. These signatures include surface shearing and rotational flows, quadrupolar geometry above the surface, central sheared 
arcades reconnecting with oppositely orientated overlying dipole fields, the formation of coronal flux ropes, and internal
reconnection which resembles the classical flare reconnection scenario.
Thus, within these simulations we have validated the use of  these proxies for flux emergence by certain CME models, and made a major step forward towards
fully self-consistent models of the buildup and eruption of magnetic energy in the corona.

\begin{acknowledgments}
\nind{Acknowledgements:}
This work has been supported by the NASA Living With a Star \& Solar and Heliospheric 
Physics programs, and the Office of Naval Research 6.1 Program. 
The simulations were performed under a grant of computer time from the DoD HPC program.
\end{acknowledgments}


\newpage

\providecommand{\noopsort}[1]{}\providecommand{\singleletter}[1]{#1}%

\end{document}